\documentclass[a4paper,preprint,superscriptaddress,nofootinbib,dvipdfmx,longbibliography]{revtex4-1}
\usepackage{graphicx}
\usepackage{bm}
\usepackage{amssymb}
\usepackage{amsmath}   
\usepackage{color}
\usepackage{cancel}
\usepackage{comment}
\usepackage{slashed}

\def\beq#1\eeq{\begin{equation}#1\end{equation}}
\def\bal#1\eal{\begin{align}#1\end{align}}

\begin{document}

\title{
Probing doubly charged scalar bosons from the doublet \\ at future high-energy colliders
}

\author{Kazuki Enomoto}
\email{kenomoto@het.phys.sci.osaka-u.ac.jp}
\affiliation{Department of Physics, Osaka University, Toyonaka, Osaka 560-0043, Japan}

\author{Shinya Kanemura}
\email{kanemu@het.phys.sci.osaka-u.ac.jp }
\affiliation{Department of Physics, Osaka University, Toyonaka, Osaka 560-0043, Japan}

\author{Kento Katayama}
\email{k\_katayama@het.phys.sci.osaka-u.ac.jp }
\affiliation{Department of Physics, Osaka University, Toyonaka, Osaka 560-0043, Japan}

\preprint{OU-HET-1079}

\begin{abstract}
The isospin doublet scalar field with hypercharge $3/2$ 
is introduced in some new physics models such as tiny neutrino masses. 
Detecting the doubly charged scalar bosons from the doublet field 
can be a good probe of such models. 
However, their collider phenomenology has not been examined sufficiently. 
We investigate collider signatures of the doubly and 
singly charged scalar bosons 
at the LHC for the high-luminosity upgraded option (HL-LHC) 
by looking at transverse mass distributions etc. 
With the appropriate kinematical cuts we demonstrate the background reduction 
 in the minimal model in the following two cases 
 depending on the mass of the scalar bosons. 
(1) The main decay mode of the singly charged scalar bosons is the tau lepton and missing (as well as charm and strange quarks). 
(2) That is into a top bottom pair.  
In the both cases, we assume that the doubly charged scalar boson is heavier than the singly charged ones. 
We conclude that the scalar doublet field with $Y=3/2$ is expected to be 
detectable at the HL-LHC unless the mass is too large. 
\end{abstract}

\maketitle

\section{Introduction}

In spite of the success of the Standard Model (SM), 
there are good reasons to regard the model as an effective theory 
around the electroweak scale, 
above which the SM should be replaced by a model of new physics beyond the SM. 
Although a Higgs particle has been discovered at the LHC~\cite{ref:Higgs_discovery}, 
the structure of the Higgs sector remains unknown. 
Indeed, the current data from the LHC can be explained in the SM. 
However, the Higgs sector in the SM causes the hierarchy problem, 
which must be solved by introducing new physics beyond the SM. 
In addition, the SM cannot explain gravity and several phenomena 
such as tiny neutrino masses, 
dark matter, baryon asymmetry of the universe, and so on. 
Clearly, extension of the SM is inevitable to explain these phenomena.   

In the SM, introduction of a single isospin doublet scalar field is just a hypothesis without any theoretical principle. 
Therefore, there is still a room to consider non-minimal shapes of the Higgs sector. 
When the above mentioned problems of the SM are considered together with such uncertainty of the Higgs sector, 
it might happen that it would be one of the natural directions to think about the possibility of extended Higgs sectors as effective theories of unknown more fundamental theories beyond the SM. 
Therefore, there have been quite a few studies on models with extended Higgs sectors both theoretically and phenomenologically.  

Additional isospin-multiplet scalar fields have often been introduced into the Higgs sector in lots of new physics models such as models of supersymmetric extensions of the SM, 
those for tiny neutrino masses~\cite{ref:Type-I_seesaw, ref:Type-II_seesaw, 
ref:Left-Right, ref:Type-III_seesaw, ref:Zee, ref:Zee_Babu, ref:Cheng_Li, ref:KNT, ref:Ma, ref:AKS, ref:Cocktail}, 
dark matter~\cite{Araki:2011hm, ref:Deshpande_Ma, ref:intert_singlet}, 
CP-violation~\cite{ref:Kobayashi_Maskawa, ref:Lee_CPviolation}, 
and the first-order phase transition~\cite{Kuzmin:1985mm, Cohen:1990it}. 
One of the typical properties in such extended Higgs sector is a prediction of existence of charged scalar states. 
Therefore, theoretical study of these charged particles and their phenomenological exploration at experiments are essentially important to test these models of new physics. 
 
There is a class of models with extended Higgs sectors in which doubly charged scalar states are predicted.  
They may be classified by the hypercharge of the isospin-multiplet scalar field in the Higgs sector; i.e. 
triplet fields with $Y=1$~\cite{ref:Type-II_seesaw, ref:Left-Right, ref:Cheng_Li}, 
doublet fields with $Y=3/2$~\cite{Aoki:2011yk, Okada:2015hia, Cheung_Okada, 
Enomoto:2019mzl, Ma:2019coj, Das:2020pai}, 
and singlet fields with $Y=2$~\cite{ref:Zee_Babu, ref:Cheng_Li, ref:Cocktail, Cheung_Okada}. 
These fields mainly enter into new physics model motivated  to explain tiny neutrino masses, 
sometimes together with dark matter and baryon asymmetry of the universe~\cite{Aoki:2011yk, Okada:2015hia, Enomoto:2019mzl, Ma:2019coj, Das:2020pai, ref:Cocktail}. 
The doubly charged scalars are also introduced in models for other motivations~\cite{Georgi:1985nv,ArkaniHamed:2002qy}. 
Collider phenomenology of these models is important to discriminate the models. 
There have also been many studies along this line~\cite{ref:Gunion, ref:triplet_pheno, Han:2007bk, Kanemura:2014goa, Aoki:2011yk, Rentala:2011mr, King:2014uha, ref:distinguish_doubly, Vega:1989tt, Han:2003wu, ref:exotic_Higgs}. 

\vspace{-5pt}

In this paper, 
we concentrate on the collider phenomenology of the model with an additional isodoublet field $\Phi$ with $Y=3/2$ at the high-luminosity-LHC (HL-LHC) with the collision energy of $\sqrt{s} = 14 ~\mathrm{TeV}$ and the integrated luminosity of $\mathcal{L} = 3000~\mathrm{fb^{-1}}$~\cite{HL-LHC}. 
Clearly, $\Phi$ cannot couple to fermions directly. 
The component fields are doubly charged scalar bosons~$\Phi^{\pm\pm}$  and singly charged ones~$\Phi^\pm$.
In order that the lightest one is able to decay into light fermions, 
we further introduce an additional doublet scalar field $\phi_2$ with the same hypercharge as of the SM one $\phi_1$, $Y=1/2$. 
Then, $Y=3/2$ component fields can decay via the mixing between two physical singly charged scalar states. 
Here, we define this model as a minimal model with doubly charged scalar bosons from the doublet. 
This minimal model has already been discussed in Ref.~\cite{Aoki:2011yk}, 
where signal events via $pp\to W^{+\ast} \to \Phi^{++} H_i^-$ 
have been analyzed, 
where $H_i^\pm$ ($i=1,2$) are mass eigenstates of singly charged scalar states. 
They have indicated that masses of all the charged states $\Phi^{\pm\pm}$ and $H_i^\pm$ may be measurable form this single process by looking at the Jacobian peaks of transverse masses of several combinations of final states etc. 
However, they have not done any analysis for backgrounds. 
In this paper, we shall investigate both signal and backgrounds for this process to see whether or not the signal can dominate the backgrounds after performing kinematical cuts at the HL-LHC. 

This paper is organized as follows. 
In Sec.~II, we introduce 
the minimal model with doubly charged scalar bosons from the doublet 
which is mentioned above, 
and give a brief comment about current constraints on the singly charged scalars from some experiments. 
In Sec.~III, 
we investigate decays of doubly and singly charged scalars 
and a production of doubly charged scalars at hadron colliders. 
In Sec.~IV, 
results of numerical evaluations for the process $pp\to W^{+\ast} \to \Phi^{++} H_i^-$ are shown. 
Final states of the process depend on mass spectrums 
of the charged scalars, and 
we investigate two scenarios with a benchmark value. 
Conclusions are given In Sec.~V. 
In Appendix~A, 
we show analytic formulae for decay rates of two-body and three-body decays of the charged scalars. 



\section{Model of the scalar field with $Y=3/2$}
 
We investigate the model whose scalar potential includes
three isodoublet scalar fields \newpage
\noindent $\phi_1$, $\phi_2$, 
and $\Phi$~\cite{Aoki:2011yk}. 
Gauge groups and fermions in the model are same with those in the SM. 
Quantum numbers of scalar fields are shown in Table~\ref{table:Scalars}. 
The hypercharge of two scalars $\phi_1$ and $\phi_2$ is $1/2$, 
and that of the other scalar $\Phi$ is $3/2$. 
In order to forbid the flavor changing neutral current (FCNC) at tree level, we impose the softly broken $Z_2$ symmetry, where $\phi_2$ and $\Phi$ have odd parity 
and $\phi_1$ has even parity~\cite{Glashow:1976nt}. 
 
 \begin{table}[h]
 \begin{center}
 \begin{tabular}{c|c|c|c|c|}
 			&	\ $SU(3)_C$\ 	&	\ $SU(2)_L$\ 	&	\ $U(1)_Y$\ 	&	\ $Z_2$\ 	\\ \hline\hline
 $\phi_1$	&	${\bf 1}$	&	${\bf 2}$	&	$1/2$		&	$+$	\\ \hline
 $\phi_2$	&	${\bf 1}$	&	${\bf 2}$	&	$1/2$		&	$-$		\\ \hline
 $\Phi$		&	${\bf 1}$	&	${\bf 2}$	&	$3/2$		&	$-$		\\ \hline
 \end{tabular}
 \caption{The list of scalar fields in the model}
\label{table:Scalars}
 \end{center}
 \end{table}
 
The scalar potential of the model is given by
\bal
V = & V_{\rm THDM} + \mu_{\Phi}^2 |\Phi|^2 + \frac{ 1 }{ 2 } \lambda_{\Phi} | \Phi |^4
	+ \sum_{i=1}^2 \rho_i  |\phi_i|^2  |\Phi|^2
	+ \sum_{i=1}^2 \sigma_i | \phi_i^\dagger \Phi|^2
\nonumber \\
	& + \Bigl\{
			\kappa (\Phi^\dagger \phi_1)(\tilde{ \phi_1}^\dagger \phi_2) 
			+ \mathrm{h.c.}
		\Bigr\}, 
\eal
where $V_{\rm THDM}$ is the scalar potential in the two Higgs doublet model 
(THDM), and it is given by
\bal
V_{\rm THDM} =& \sum_{i = 1}^2 \mu_i^2 |\phi_i|^2
			+\Bigl(
				\mu_3^2 \phi_1^\dagger \phi_2 + \mathrm{h.c.}
			\Bigr)
			+ \sum_{i=1}^2 \frac{ 1 }{ 2 } \lambda_i |\phi_i|^4
			+ \lambda_3 |\phi_1|^2 |\phi_2|^2
			+ \lambda_4 | \phi_1^\dagger \phi_2 |^2
\nonumber \\ 
			& + \frac{ 1 }{ 2 } 
			\Bigl\{
				\lambda_5 ( \phi_1^\dagger \phi_2 )^2 
				+ \mathrm{h.c.}
			\Bigr\}.
\eal
The $Z_2$ symmetry is softly broken by the terms of $\mu_3^2 \phi_1^\dagger \phi_2$ and its hermitian conjugate. 
Three coupling constants $\mu_3$, $\lambda_5$ and $\kappa$ can be complex number generally. 
After redefinition of phases of scalar fields, either $\mu_3$ or $\lambda_5$ remains as the physical CP-violating parameter. 
In this paper, we assume that this CP-violating phase is zero and all coupling constants are real for simplicity.  

Component fields of the doublet fields are defined as follows. 
\bal
\phi_i = 
\begin{pmatrix}
\omega_i^+ \\
\frac{ 1 }{ \sqrt{2} } ( v_i + h_i + i z_i ) \\
\end{pmatrix}
, \quad 
\Phi = 
\begin{pmatrix}
\Phi^{++} \\
\Phi^+ \\
\end{pmatrix}, 
\eal
where $i=1,2$. 
The fields $\phi_1$ and $\phi_2$ obtain the vacuum expectation values (VEVs) $v_1/\sqrt{2}$ and $v_2/\sqrt{2}$, respectively. 
These VEVs are described by $v \equiv \sqrt{v_1^2 + v_2^2} \simeq 246\ \mathrm{GeV}$ and $\tan \beta \equiv v_2 / v_1$.  
On the other hand, the doublet $\Phi$ cannot have a VEV without violating electromagnetic charges spontaneously.  

Mass terms for the neutral scalars $h_i$ and $z_i$ are generated by $V_{\rm THDM}$. 
Thus, mass eigenstates of the neutral scalars are defined in the same way with those in the THDM 
(See, for example, Ref.~\cite{Branco:2011iw}). 
Mass eigenstates $h$, $H$, $A$, and $z$ are defined as 
\bal
\begin{pmatrix}
H \\ 
h \\
\end{pmatrix}
= R(\alpha)
\begin{pmatrix}
h_1 \\
h_2 \\
\end{pmatrix}
, \quad 
\begin{pmatrix}
z \\
A \\
\end{pmatrix}
= R(\beta)
\begin{pmatrix}
z_1 \\
z_2 \\
\end{pmatrix}
, 
\eal
where $\alpha$ and $\beta$ ($= \mathrm{Tan}^{-1}( v_2 / v_1)$) are mixing angles, 
and $R(\theta)$ is the two-by-two rotation matrix for the angle $\theta$, which is given by
\bal
R(\theta) = 
\begin{pmatrix}
\cos \theta & \sin \theta \\
- \sin \theta & \cos \theta \\
\end{pmatrix}. 
\eal
The scalar $z$ is the Nambu-Goldstone (NG) boson, 
and it is absorbed into the longitudinal component of $Z$ boson. 
Thus, the physical neutral scalars are $h$, $H$, and $A$. 
For simplicity, we assume that $\sin(\beta - \alpha) = 1$ so that $h$ is the SM-like Higgs boson. 

On the other hand, the mass eigenstates of singly charged scalars are different from 
those in the THDM, 
because the field $\Phi^\pm$ is mixed with $\omega_1^\pm$ and $\omega_2^\pm$. 
The singly charged mass eigenstates $\omega^\pm$, $H_1^\pm$, and $H_2^\pm$ are defined as 
\bal
\begin{pmatrix}
\omega^\pm \\
H_1^\pm \\
H_2^\pm \\
\end{pmatrix}
=
\begin{pmatrix}
1 & 0 & 0 \\
0 & \cos \chi & \sin \chi \\
0 & -\sin \chi & \cos \chi \\
\end{pmatrix}
\begin{pmatrix}
\cos \beta & \sin \beta & 0 \\
- \sin \beta & \cos \beta & 0 \\
0 & 0 & 1 \\
\end{pmatrix}
\begin{pmatrix}
\omega_1^\pm \\
\omega_2^\pm \\
\Phi^\pm \\
\end{pmatrix}.
\eal
The scalar $\omega^\pm$ is the NG boson, and it is absorbed into the longitudinal component of $W^\pm$ boson. 
Thus, there are two physical singly charged scalars $H_1^\pm$ and $H_2^\pm$. 
The doubly charged scalar $\Phi^{\pm\pm}$ is mass eigenstate without mixing. 

The doublet $\Phi$ does not have the Yukawa interaction with the SM fermions 
because of its hypercharge.\footnote{
If we consider higher dimensional operators, 
interactions between $\Phi$ and leptons are allowed~\cite{Rentala:2011mr}.} 
Therefore, Yukawa interactions in the model is same with those in the THDM. 
They are divided into four types according to the $Z_2$ parities of each fermion 
(Type-I, II, X, and Y~\cite{Aoki:2009ha}). 
In the following, we consider the Type-I Yukawa interaction where all left-handed fermions have even parity, and all right-handed ones have odd-parity. 
The type-I Yukawa interaction is given by
\bal
\label{eq:Yukawa}
\mathcal{L}_{Yukawa}
= - \sum_{i, j = 1}^3
\biggl\{
	(Y_u)_{ij} \overline{Q}_{iL} \tilde{\phi}_2 u_{jR}^{}
	+ (Y_d)_{ij} \overline{Q_{iL}} \phi_2 d_{jR}^{}
	+ (Y_\ell)_{ij} \overline{ L_{iL} } \phi_2 \ell_{jR}^{}
\biggr\} + \mathrm{h.c.}, 
\eal
where $Q_{iL}$ $(L_{iL})$ is the left-handed quark (lepton) doublet, 
and $u_{jR}^{}$, $d_{jR}$, and $\ell_{jR}$ are the right-handed 
up-type quark, down-type quark and charged lepton fields, respectively. 
The Yukawa interaction of the singly charged scalars are given by 
\beq
\label{eq:charged_Yukawa}
- \frac{ \sqrt{2} }{ v } \cot \beta \sum_{i,j =1}^3
\biggl\{
	V_{u_i d_j} \overline{u_i} 
	\Bigl( m_{u_i} P_L + m_{d_j} P_R \Bigr) d_{j}
	+ \delta_{ij} m_{\ell_i} \overline{ \nu_i } P_L \ell_i 
\biggr\}
\Bigl( \cos \chi H_1^+ - \sin \chi H_2^+ \Bigr) + \mathrm{h.c.}, 
\eeq
where $V_{u_i d_j}$ is the $(u_i ,d_j)$ element of the Cabibbo-Kobayashi-Maskawa (CKM) matrix~\cite{Cabibbo:1963yz, ref:Kobayashi_Maskawa}, 
$\delta_{ij}$ is the Kroneker delta, 
and $P_L$ ($P_R$) is the chirality projection operator for left-handed (right-handed) chirality. 
In addition, $(u_1, u_2, u_3) = (u, c, t)$ are the up-type quarks, 
$(d_1, d_2, d_3) = (d, s, b)$ are the down-type quarks, 
$(\ell_1, \ell_2, \ell_3) =  (e, \mu, \tau)$ are the charged leptons, 
and $(\nu_1, \nu_2, \nu_3) = (\nu_e, \nu_\mu, \nu_\tau)$ are the neutrinos. 
The symbols $m_{u_i}$, $m_{d_i}$, and $m_{\ell_i}$ are the masses for 
$u_i$, $d_i$, and $\ell_i$, respectively. 
In the following discussions, 
we neglect non-diagonal terms of the CKM matrix.

Finally, we discuss constraints on some parameters in the model from various experiments. 
If the coupling constant $\kappa$ in the scalar potential is zero, the model have a new discrete $Z_2$ symmetry where the doublet $\Phi$ is odd  and all other fields are even. 
This $Z_2$ symmetry stabilizes $\Phi^{\pm\pm}$ or $\Phi^\pm$, and their masses and interactions are strongly constrained. 
Thus, $\kappa \neq 0$ is preferred, and it means that $\sin \chi \neq 0$. 
In this paper, we assume that $\chi = \pi / 4$ just for simplicity. 
Since the charged scalars $H_1^\pm$ and $H_2^\pm$ have Type-I Yukawa interaction, 
it is expected that the constraints on $H_1^\pm$ and $H_2^\pm$ are almost same with 
those on the charged Higgs boson in the Type-I THDM 
and the difference is caused by the factor $\sin \chi$ or $\cos \chi$ 
in Eq.~(\ref{eq:charged_Yukawa}).  
In the case where $\sin \chi = \cos \chi = 1 / \sqrt{2}$, 
the constraints are as follows. 
For $\tan \beta \lesssim 1.4$, the lower bound on the masses of $H_1^\pm$ and $H_2^\pm$ are given by flavor experiments. 
This lower bound depends on the value of $\tan \beta$, 
and it is about $400~\mathrm{GeV}$ for $\tan \beta = 1$~\cite{Enomoto:2015wbn, Arbey:2017gmh,  Haller:2018nnx}. 
In the region that $1.4 \lesssim \tan \beta \lesssim 5.7$, 
the lower bound on the mass is given by 
the search for the decay of the top quark into the bottom quark and the singly charged scalar at the LHC Run-I.  
This lower bound is about $170~\mathrm{GeV}$~\cite{Arbey:2017gmh, Aiko:2020ksl}. 
For $\tan \beta \gtrsim 5.7$, the direct search at LEP gives the lower bound on the mass.  It is about $80~\mathrm{GeV}$~\cite{Abbiendi:2013hk}. 
From Eq.~(\ref{eq:charged_Yukawa}), it is obvious that 
if we think the case where $|\sin \chi| > | \cos \chi |$, ($|\sin \chi| < | \cos \chi |$)
the constraints on $H_1^\pm$ ($H_2^\pm$) are relaxed, 
and those on $H_2$ ($H_1^\pm$) become more stringent.



\section{Production and decays of charged scalar states}

In this section, we investigate the decay of the new charged scalars and the production of the doubly charged scalar at hadron colliders. 
In the following discussion, 
we assume that $\Phi^{\pm\pm}$, $H$, and $A$ are heavier than ${H_1}^\pm$ and ${H_2}^\pm$. 
Then, $H_{1,2}^\pm$ cannot decay into $\Phi^{\pm\pm}$, $H$, and $A$. 
In addition, the masses of $H_1^\pm$, $H_2^\pm$, and $\Phi^{\pm\pm}$ are denoted by 
$m_{H_1}^{}$ $m_{H_2}^{}$, and $m_{\Phi}^{}$, respectively. 

\subsection{Decays of charged scalar sates}

First, we discuss the decays of the singly charged scalars $H_1^\pm$ and $H_2^\pm$. 
They decay into the SM fermions via Yukawa interaction in Eq.~(\ref{eq:charged_Yukawa}). 
Since they are lighter than $\Phi^{\pm\pm}$, $H$, and~$A$, 
their decays into $\Phi^{\pm\pm} W^{\mp(\ast)}$, $H W^{\pm(\ast)}$, 
and $A W^{\pm(\ast)}$ are prohibited. 
On the other hand, 
the decay of the heavier singly charged scalars into the lighter one and $Z^{(\ast)}$ is allowed, 
and it is generated via the gauge interaction. 
In the following, 
we assume that $H_2^\pm$ is heavier than $H_1^\pm$ ($m_{H_2} > m_{H_1}$). 

\begin{figure}[h]
\begin{center}
\includegraphics[width=80mm]{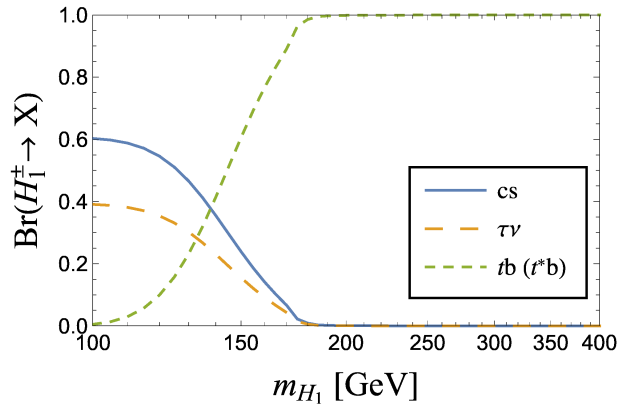}
\caption{The branching ratio of $H_1^\pm$.}
\label{fig:Decay_H1}
\end{center}
\end{figure}

In Fig.~\ref{fig:Decay_H1}, 
the branching ratio for each decay channel of $H_1^\pm$ is shown. 
Since we assume that $H_1^\pm$ is lighter than $H_2^\pm$, 
it decays via the Yukawa interaction~\cite{Aoki:2009ha}\footnote{
In this paper, we neglect the effects of one-loop induced decays 
$H_i^\pm \to W^\pm \gamma$
and $H_i\pm \to W^\pm Z$~\cite{CapdequiPeyranere:1990qk}.}. 
In the region where $m_{H_1}^{} \lesssim 140~\mathrm{GeV}$, 
the decay into $cs$ and that into $\tau \nu$ are dominant. 
When we consider a little heavier $H_1^\pm$, which are in the mass region 
between $140~\mathrm{GeV}$ and $m_t + m_b \simeq 180~\mathrm{GeV}$, 
the branching ratio for $H_{1,2}^\pm \to t^\ast b \to W^\pm b \overline{b}$ is dominant~\cite{Ma:1997up}.\footnote{
In Ref~\cite{Ma:1997up}, Type-II Yukawa interaction is investigated, and the condition $\tan \beta \lesssim 1$ is needed to make the decay $H_{1,2}^\pm \to t^\ast b$ dominant. 
In our case (Type-I), this condition is not necessary
because all fermions couple to $\phi_2$ universally.}
In the mass region $m_t + m_b < m_{H_1}^{}$, 
the branching ratio for $H_1^\pm \to tb$ is almost $100~\%$. 
The decays into $cs$, $\tau \nu$, and $t^{(\ast)} b$ are all induced by 
the Yukawa interaction. 
Since we consider the Type-I Yukawa interaction,  
the dependence on $\tan \beta$ of each decay channel is same. 
Thus, the branching ratio in Fig.~\ref{fig:Decay_H1} hardly depends on the value of $\tan \beta$.
Analytic formulae of decay rates for each decay channel are shown in Appendix~\ref{sec:Decay_of_H}. 

\begin{figure}[h]
\begin{tabular}{c}
	\begin{minipage}[t]{0.5\hsize}
		\begin{center}
		\includegraphics[width=80mm]{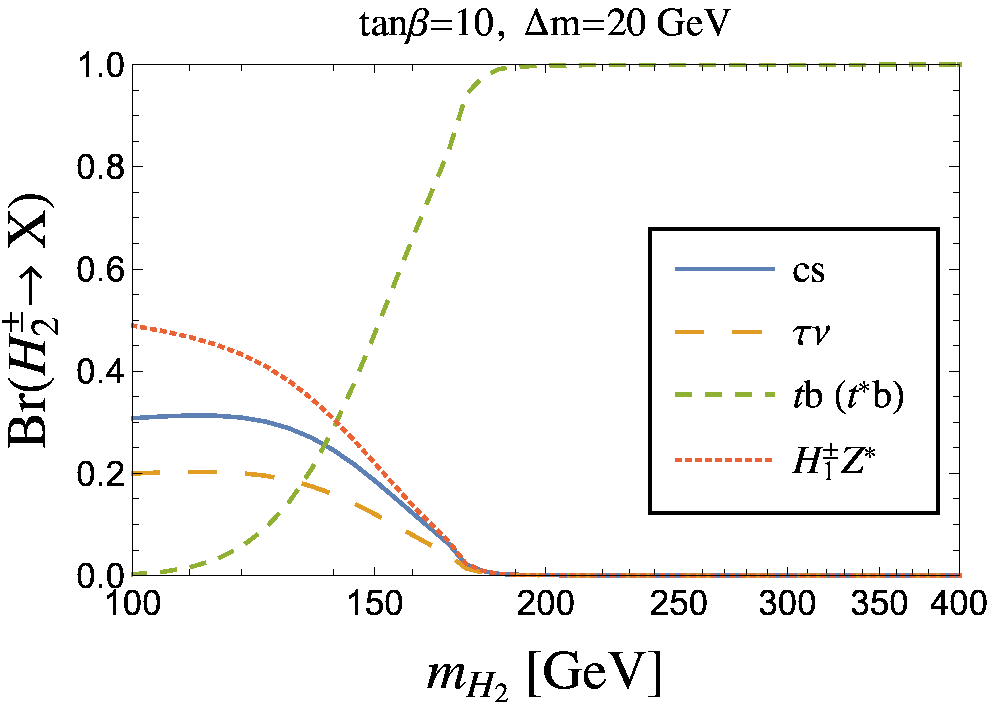}
		\end{center}
	\end{minipage}
	\begin{minipage}[t]{0.5\hsize}
		\begin{center}
		\includegraphics[width=80mm]{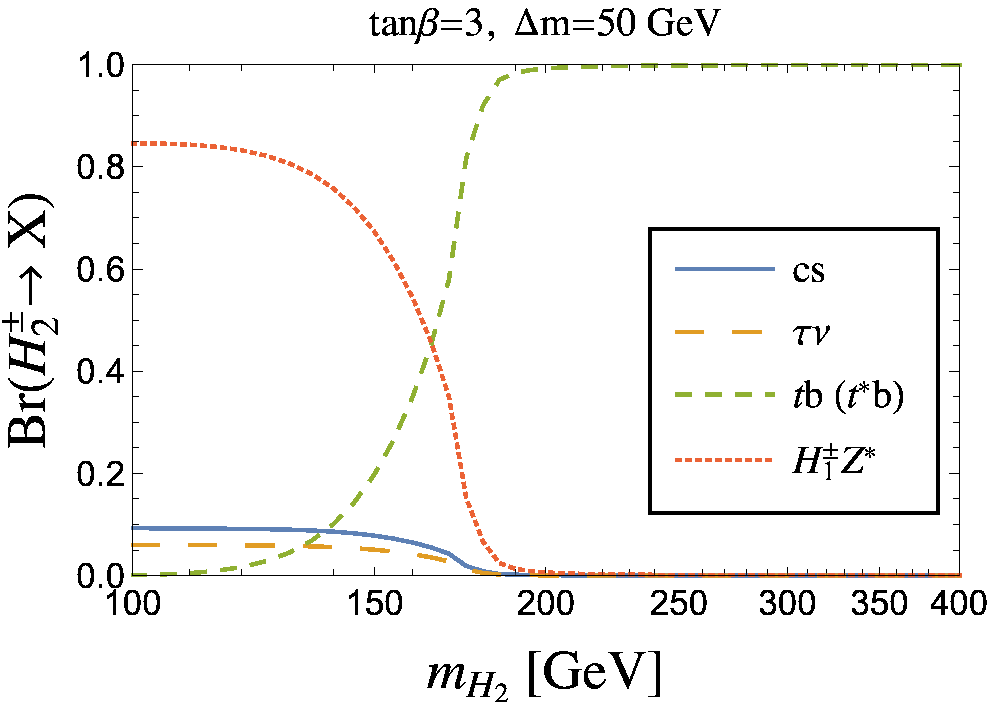}
		\end{center}
	\end{minipage}
\end{tabular}
\caption{The branching ratio of $H_2^\pm$.
		In the left figure, we assume that 
		$\Delta m(\equiv m_{H_2}^{} - m_{H_1}^{}) = 20~\mathrm{GeV}$ and $\tan \beta = 10$. 
		In the right figure, we assume that 
		$\Delta m = 50~\mathrm{GeV}$ and $\tan \beta = 3$}
		\label{fig:Decay_H2}
\end{figure}

The singly charged scalar $H_2^\pm$ also decays into the SM fermions via the Yukawa interaction. 
In addition, 
$H_2^\pm \to H_1^\pm Z^{(\ast)}$ is allowed. 
In Fig.~\ref{fig:Decay_H2}, 
the branching ratios of $H_2^\pm$ in two cases are shown. 
The left figure of Fig.~\ref{fig:Decay_H2} is 
for  $\tan \beta = 10$ and 
$\Delta m (\equiv m_{H_2} - m_{H_1}) = 20~\mathrm{GeV}$. 
In the small mass region, 
the decay $H_2^\pm \to H_1^\pm Z^\ast$ is dominant. 
In the region where $m_{H_2}^{} \gtrsim 140~\mathrm{GeV}$, 
the decay $H_2^\pm \to t^{(\ast)} b$ becomes dominant, 
and the branching ratio for $H_2^\pm \to tb$ is almost $100~\%$ 
for $m_{H_2}^{} \gtrsim 180~\mathrm{GeV}$. 
If we consider smaller $\tan \beta$, 
the decays via Yukawa interaction are enhanced because the Yukawa interaction is proportional to $\cot \beta$. (See Eq.~(\ref{eq:charged_Yukawa}).) 
Thus, he branching ratio for $H_2^\pm \to H_1^\pm Z^\ast$ decreases. 

The right figure of Fig.~\ref{fig:Decay_H2} is for the case where 
$\tan \beta = 3$ and $\Delta m = 50~\mathrm{GeV}$.
In the small mass region, 
the branching ratio for $H_2^\pm \to H_1^\pm Z^\ast$ is about $80~\%$, 
and those for other decay channels are negligible small. 
However, in the mass region where $m_{H_2}^{} \gtrsim 180~\mathrm{GeV}$, 
$H_2^\pm \to H_1^\pm Z^\ast$ become negligible small, 
and the branching ratio for $H_2^\pm \to tb$ is almost $100~\%$. 
If we consider larger $\tan \beta$, 
the decays via the Yukawa interaction is suppressed, 
and the branching ratio for $H_2^\pm \to H_1^\pm Z^\ast$ 
increases. 
Thus, the crossing point of 
the branching ratio for $H_2^\pm \to tb (t^\ast b)$ 
and that for $H_2^\pm \to H_1^\pm Z^\ast$ 
move to the point at heavier $m_{H_2}^{}$. 
Analytic formulae of decay rates for each decay channel are shown in Appendix~\ref{sec:Decay_of_H}. 

Next, we discuss the decay of the doubly charged scalar $\Phi^{\pm\pm}$. 
The doubly charged scalar $\Phi^{\pm\pm}$ does not couple to fermions via Yukawa interaction\footnote{
This is different from doubly charged Higgs boson in the triplet model 
in which dilepton decays of doubly charged Higgs bosons are important signature to test the model~\cite{Han:2003wu}.}. Therefore, it decays via the weak 
\newpage
\noindent gauge interaction\footnote{
In triplet Higgs models, if the VEV of the triplet field is small enough the main decay mode of the doubly charged Higgs boson is the diboson decay~\cite{Kanemura:2014goa}.  
On the other hand, in our model, such a decay mode does not exist at tree level.}. 
We consider the following three cases.

\begin{figure}[h]
\begin{tabular}{c}
	\begin{minipage}[t]{0.5\hsize}
		\begin{center}
		\includegraphics[width=80mm]{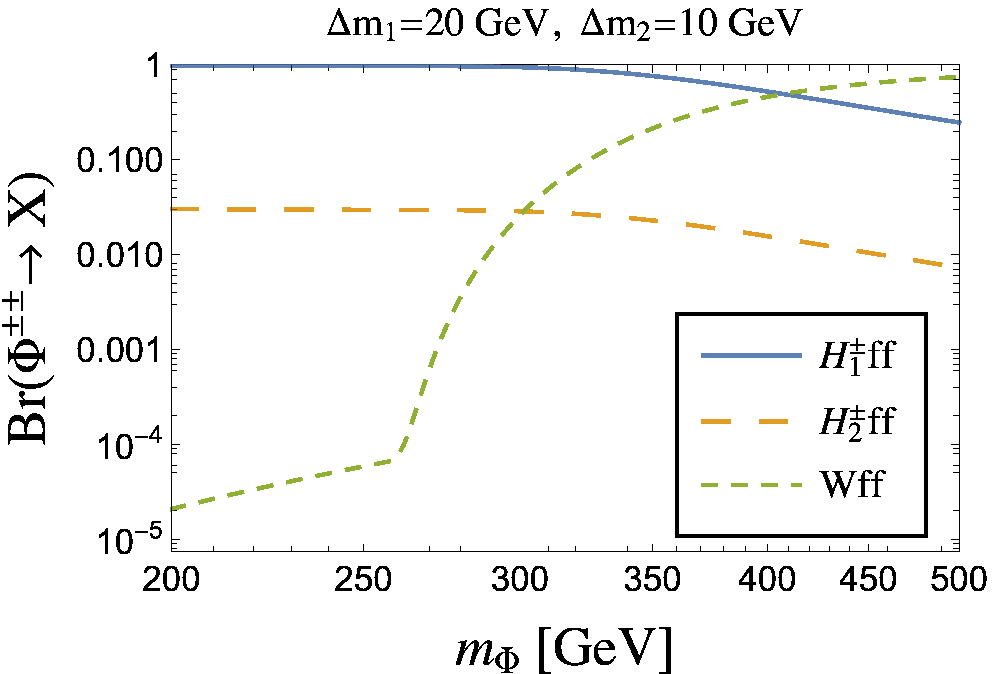}
		\end{center}
	\end{minipage}
	\begin{minipage}[t]{0.5\hsize}
		\begin{center}
		\includegraphics[width=80mm]{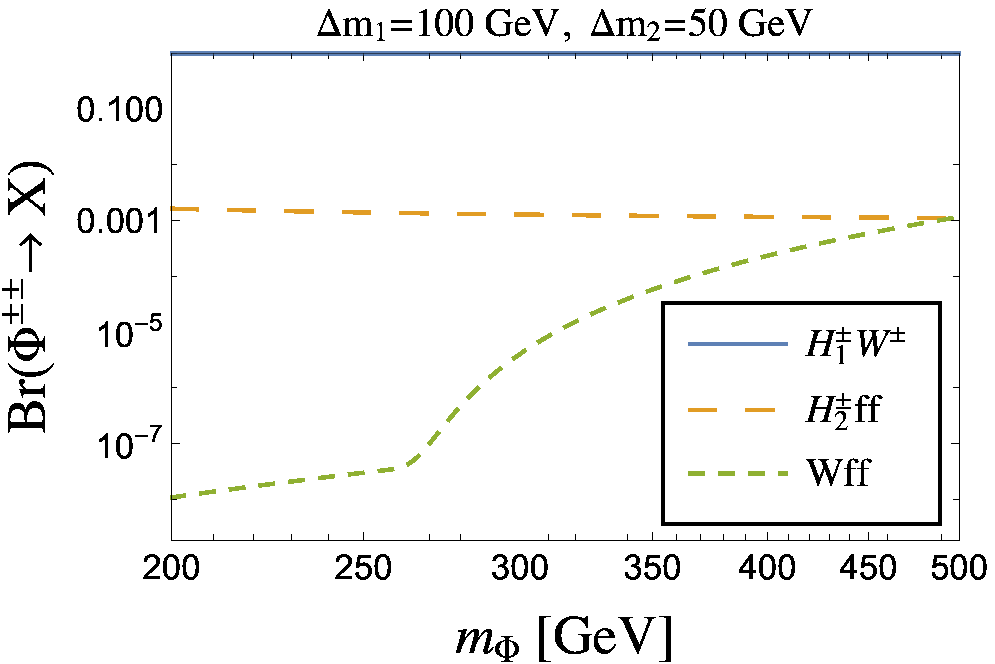}
		\end{center}
	\end{minipage}
\end{tabular}
\begin{center}
\includegraphics[width=80mm]{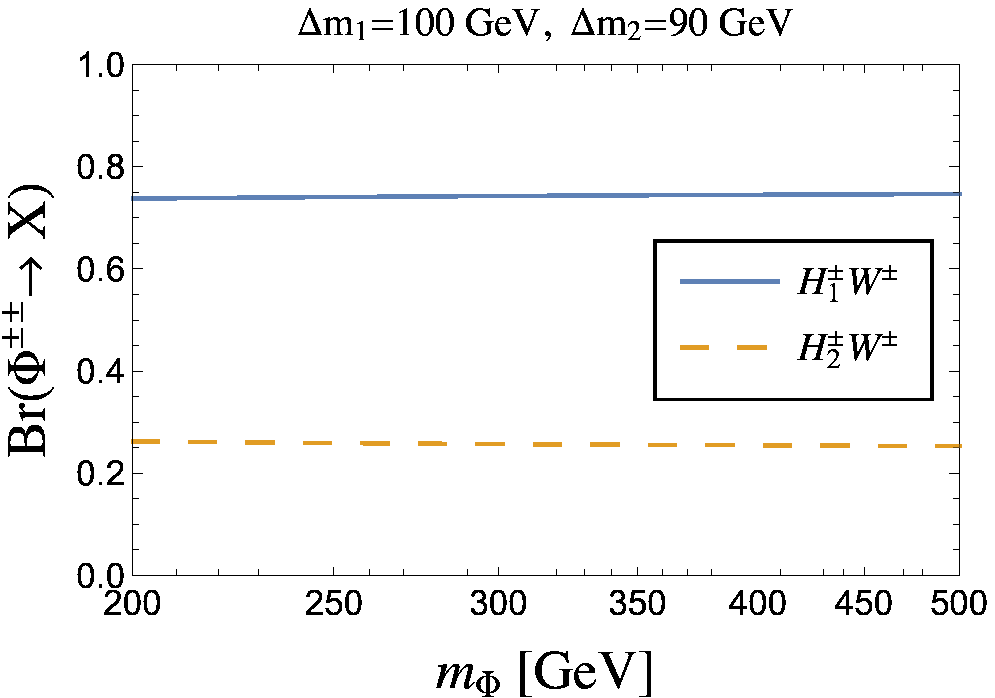}
\end{center}
\caption{The branching ratios of the decay of $\Phi^{\pm\pm}$. 
		The upper lift (right) afigure is those in the case that 
		$\Delta m_1^{}(\equiv m_{\Phi}^{} - m_{H_1}^{}) = 20~\mathrm{GeV}$ ($100~\mathrm{GeV}$) 
		and $\Delta m_2^{}(\equiv m_{\Phi}^{} - m_{H_2}^{}) = 10~\mathrm{GeV}$ ($50~\mathrm{GeV}$). 
		The bottom one corresponds to the case that 
		$\Delta m_1 = 100~\mathrm{GeV}$ and 
		$\Delta m_2 = 90~\mathrm{GeV}$.}
\label{fig:Decay_Phi}
\end{figure}

First, the case where $\Delta m_1(\equiv m_{\Phi}^{} - m_{H_1}^{}) < 80~\mathrm{GeV}$ and 
$\Delta m_2 (\equiv m_{\Phi}^{} - m_{H_2}^{})  < 80~\mathrm{GeV}$ is considered.  
In this case, 
$\Phi^{\pm\pm}$ cannot decay into the on-shell $H_{1,2}^\pm$, 
and three-body decays are dominant. 
In the upper left figure of Fig.~\ref{fig:Decay_Phi}, 
the branching ratio of $\Phi^{\pm\pm}$ in this case is shown. 
We assume that $\tan \beta = 3$, $\Delta m_1 < 20~\mathrm{GeV}$, 
$\Delta m_2 < 10~\mathrm{GeV}$. 
In the small mass region, 
$\Phi^{\pm\pm} \to H_1^\pm ff$ is dominant. 
With increasing of $m_{\Phi}$, 
the masses of $H_{1,2}^\pm$ also increase because the mass differences between them are fixed. 
Thus, the branching ratio for $\Phi^{\pm\pm} \to W^\pm ff$ is dominant in the large mass region. 
At the point $m_{\Phi} \simeq 260~\mathrm{GeV}$, 
the branching ratio for $\Phi^{\pm\pm} \to W^\pm ff$ changes rapidly. 
It is because that at this point, 
the decay channel $\Phi^{\pm\pm} \to W^\pm t b$ is open. 
If we consider the large $\tan \beta$, 
the decay rates of $\Phi^{\pm\pm} \to W^\mp ff$ becomes small 
because this process includes $H_{1,2}^{\pm \ast} \to f f$ via Yukawa interaction which is proportional to $\cot \beta$. 
However, the decays $\Phi^{\pm\pm} \to H_{1,2}^\pm ff$ are generated via only the gauge interaction. 
Thus, for $\tan \beta \gtrsim 3$, 
the branching ratio for $\Phi^{\pm\pm} \to W^\pm ff$ becomes small. 

Second, the case where $\Delta m_1 > 80~\mathrm{GeV}$ and $\Delta m_2 < 80~\mathrm{GeV}$ is considered. 
In this case, $\Phi^{\pm\pm} \to H_1^\pm W^\pm$ is allowed while 
$\Phi^{\pm\pm} \to H_2^\pm W^\pm$ is prohibited. 
In the upper right figure of Fig.~\ref{fig:Decay_Phi}, 
the branching ratio of $\Phi^{\pm\pm}$ in this case is shown. 
We assume that $\tan \beta = 3$, 
$\Delta m_1 < 100~\mathrm{GeV}$, 
$\Delta m_2 < 50~\mathrm{GeV}$. 
In all mass region displayed in the figure, 
the branching ratio for $\Phi^{\pm\pm} \to H_1^\pm W^\pm$ are almost $100~\%$, 
and those for other channels are at most about $0.1~\%$. 
At the point $m_{\Phi} \simeq 260~\mathrm{GeV}$, 
the branching ratio for $\Phi^{\pm\pm} \to W^\pm ff$ changes rapidly. 
It is because that at this point, 
the decay channel $\Phi^{\pm\pm} \to W^\pm t b$ is open. 

Third, the case where $\Delta m_1 > 80~\mathrm{GeV}$ and 
$\Delta m_2  > 80~\mathrm{GeV}$ is considered.  
and both of $\Phi^{\pm\pm} \to H_{1,2}^\pm W^\pm$ are allowed. 
In the lower figure of Fig.~\ref{fig:Decay_Phi}, 
the branching ratio in this case is shown. 
We assume that $\tan \beta = 3$, 
$\Delta m_1 = 100~\mathrm{GeV}$, 
$\Delta m_2 = 90~\mathrm{GeV}$. 
In all mass region displayed in the figure, 
the branching ratio does not change because 
the mass differences between $\Phi^{\pm\pm}$ and $H_{1,2}^\pm$ are 
fixed. 
The branching ratio for $\Phi^{\pm\pm} \to H_1^\pm W^\pm$ is about $75~\%$, 
and that for $\Phi^{\pm\pm} \to H_2^\pm W^\pm$ is about $25~\%$. 
These decays are generated via only the gauge interaction. 
Thus, the branching ratios of them do not depend on $\tan \beta$, 
and they are determined by only the mass differences between $\Phi^{\pm\pm}$ and $m_{H_{1,2}}^{}$. 

\subsection{Production of $\Phi^{\pm\pm}$ at hadron colliders}

We here discuss the production of the doubly charged scalar $\Phi^{\pm\pm}$. 
In our model, production processes of charged scalar states are 
$pp\to W^{+\ast} \to H_i^+ A (H)$, 
$pp\to Z^\ast (\gamma) \to H_i^+ H_i^-$,  
$pp\to W^{+\ast} \to \Phi^{++} H_i^-$, 
and $pp\to Z^\ast (\gamma) \to \Phi^{++} \Phi^{--}$. 
In the THDM, 
the first and second processes (the singly charged scalar production) 
can also occur~\cite{ref:pair_production, ref:Associated_production} 
However, doubly charged scalar bosons are not included in the THDM\footnote{
In the THDM, and also in our model with the $Y=3/2$ doublet, there are also single production processes of singly charged Higgs bosons such as 
$g b \to t H^\pm$~\cite{Gunion:1986pe}, 
$q b \to q^\prime b H^\pm$~\cite{Moretti:1996ra}, 
$b \overline{b} \to W^\pm H^\mp$~\cite{ref:WH_associated, Asakawa:2005nx}, 
$gg \to W^\pm H^\mp$~\cite{Asakawa:2005nx, ref:gluon_fusion}, etc. (See also Ref.~\cite{Akeroyd:2016ymd}.)
In this paper, we do not consider these processes and concentrate only on the processes $pp \to W^{+\ast} \to \Phi^{++} H_i^-$.}.
In the model with the isospin triplet scalar 
with $Y=1$~\cite{ref:Type-II_seesaw, ref:Left-Right, ref:Cheng_Li, ArkaniHamed:2002qy, Georgi:1985nv}, 
all of these production processes can appear. 
However, the main decay mode of doubly charged scalar is different from our model. 
In the triplet model, 
the doubly charged scalar from the triplet mainly decays into dilepton~\cite{Han:2003wu} or diboson~\cite{Kanemura:2014goa}. 
In our model, on the other hand, 
$\Phi^{\pm\pm}$ mainly decays into the singly charged scalar and $W$ boson.

In this paper, we investigate the associated production $pp \to W^{+\ast} \to \Phi^{++} H_i^-$ $(i=1,2)$. In this process, 
informations on masses of all the charged states $\Phi^{\pm\pm}$ and $H_i^\pm$ appear in the Jacobian peaks of transverse masses of several combinations of final states~\cite{Aoki:2011yk}. 
Pair productions are also important in searching for $\Phi^{\pm\pm}$ and $H_i^\pm$, 
however we focus on the associated production in this paper. 
The parton-level cross section of the process 
$q \overline{q^\prime} \to W^{+ \ast } \to \Phi^{++} H_i^-$ 
($ i = 1,2$) is given by 
\beq
\label{eq:production_of_Phi}
\sigma_i 
= \frac{ G_F^2 m_W^4 |V_{q q^\prime}|^2 \chi_i^2 }{ 12 \pi s^2 (s - m_W^2 )^2 }
\Bigl[
	m_{H_i^\pm}^4 
	+ ( s - m_{\Phi^{\pm\pm}}^2 )^2 
	- 2 m_{H_i^\pm}^2 ( s + m_{\Phi^{\pm\pm}}^2 )
\Bigr]^{3/2}, 
\eeq
where $s$ is the square of the center-of-mass energy, $G_F$ is the Fermi coupling constant, 
and $V_{q q^\prime}$ is the $(q, q^\prime)$ element of CKM matrix. 
In addition, $\chi_i$ in Eq.~(\ref{eq:production_of_Phi}) is defined as 
\beq
\label{eq:chi_i}
\chi_1 = \sin \chi, 
\quad
\chi_2 = \cos \chi. 
\eeq

\begin{figure}[h]
\begin{center}
\includegraphics[width=100mm]{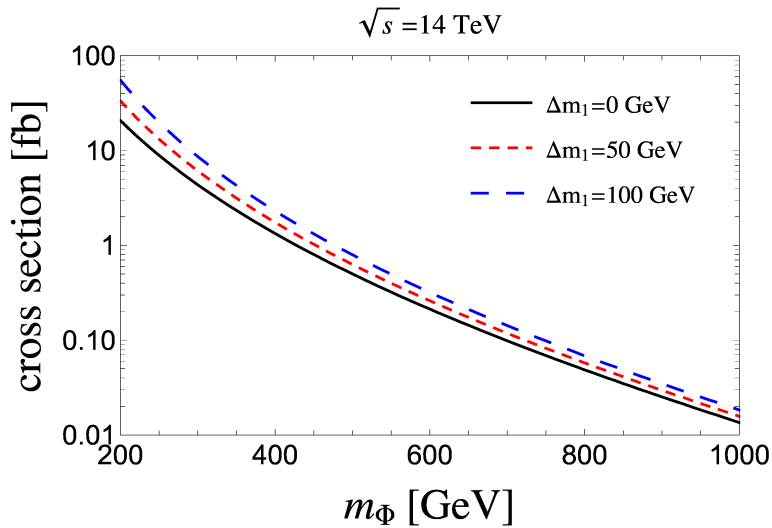}
\caption{The cross section for $pp\to W^{+\ast} \to \Phi^{++} H_1^-$, where $\sqrt{s} = 14~\mathrm{TeV}$ and $\chi = \pi/4$. 
The black, red, blue lines are those in the case that 
$\Delta m_1 ( \equiv m_{\Phi}^{} - m_{H_1}^{}) = 0$, $50$, and $100~\mathrm{GeV}$, respectively.} 
\label{fig:Production_Phi}
\end{center}
\end{figure}

In Fig.~\ref{fig:Production_Phi}, 
we show the cross section for $pp \to W^{+\ast} \to \Phi^{++} H_1^-$ 
in the case that $\sqrt{s} = 14~\mathrm{TeV}$ and 
$\chi = \pi / 4$. 
The cross section is calculated by using M{\small AD}G{\small RAPH}5\_{\small A}MC@NLO~\cite{Alwall:2014hca} and FeynRules~\cite{FeynRules}. 
The black, red, blue lines are those in the case that 
$\Delta m_1 = 0$, $50$, and $100~\mathrm{GeV}$, respectively. 
The results in Fig.~\ref{fig:Production_Phi} do not depend on the value of $\tan \beta$. 
At the HL-LHC ($\sqrt{s} = 14~\mathrm{TeV}$ and $\mathcal{L} = 3000~\mathrm{fb^{-1}}$), 
about the $6 \times 10^4$ doubly charged scalars are expected 
to be generated in the case that 
$m_{\Phi}^{} = 200~\mathrm{GeV}$ and 
$\Delta m_1 = 50~\mathrm{GeV}$. 
If $\Phi^{\pm\pm}$ is heavier, 
the cross section decreases, 
and about the $300$ doubly charged scalars are expected to be generated 
at the HL-LHC in the case that 
$m_{\Phi}^{} = 800~\mathrm{GeV}$. 
The cross section increases with increasing of the mass difference $\Delta m_1$. 
Since we assume that $\chi = \pi / 4$, 
the cross section of the process $pp \to W^{+\ast} \to \Phi^{++} H_2^-$ is same with that in Fig.~\ref{fig:Production_Phi} 
if $m_{H_2}^{} = m_{H_1}^{}$. 
If we consider the case that $| \sin \chi | > |\cos \chi|$ 
($|\cos \chi | > | \sin \chi |$), 
the cross section of $pp \to W^{+\ast} \to \Phi^{++} H_1^-$ 
become larger (smaller) than 
that of $pp \to W^{+\ast} \to \Phi^{++} H_2^-$ 
even if $m_{H_2}^{} = m_{H_1}^{}$. 

\section{Signal and backgrounds at HL-LHC}

In this section, we investigate the detectability of the process $pp \to W^{+\ast} \to \Phi^{++} H_i^-$ ($i=1,2$) 
in two benchmark scenarios. 
In the first scenario (Scenario-I), 
the masses of $H_1^\pm$ and $H_2^\pm$ are set to be $100~\mathrm{GeV}$ and $120~\mathrm{GeV}$, 
so that they cannot decay into $tb$. 
In this case, their masses are so small that the branching ratio for three body decay 
$H_{1,2}^\pm \to W^\pm b \overline{b}$ is less than $5~\%$ approximately. 
Thus, their main decay modes are $H_{1,2}^\pm \to cs$ and $H_{1,2}^\pm \to \tau \nu$. 
In the second scenario (Scenario-II), 
masses of $H_1^\pm$ and $H_2^\pm$ are set to be 
$200~\mathrm{GeV}$ and $250~\mathrm{GeV}$, 
and they predominantly decay into $tb$ with the branching ratio to be almost $100~\%$. 

In our analysis below, 
we assume the collider performance at HL-LHC as follows~\cite{HL-LHC}. 
\beq
\label{eq:HL-LHC}
\sqrt{s} = 14~\mathrm{TeV}, \quad 
\mathcal{L} = 3000~\mathrm{fb^{-1}},
\eeq 
where $\sqrt{s}$ is the center-of-mass energy and $\mathcal{L}$ is the integrated luminosity. 
Furthermore, we use the following kinematical cuts (basic cuts) for the signal event~\cite{Alwall:2014hca}; 
\beq
\label{eq:basic_cut}
\begin{array}{l}
p_T^j > 20~\mathrm{GeV}, \quad 
p_T^\ell > 10~\mathrm{GeV}, \quad
| \eta_j | < 5, \quad
| \eta_\ell | < 2.5,  \\
\Delta R_{jj} > 0.4, \quad
\Delta R_{\ell j} > 0.4, \quad
\Delta R_{\ell \ell} > 0.4,  \\
\end{array}
\eeq
where $p_T^j$ ($p_T^\ell$) and $\eta_j$ ($\eta_\ell$) 
are the transverse momentum 
and the pseudo rapidity of jets (charged leptons), respectively, 
and $\Delta R_{jj}$, $\Delta R_{\ell j}$, and $\Delta R_{\ell \ell}$ 
in Eq.~(\ref{eq:basic_cut}) are the angular distances between two jets, 
charged leptons and jets, and two charged leptons, respectively. 

\subsection{Scenario-I}

\begin{figure}[h]
\begin{center}
\includegraphics[width=80mm]{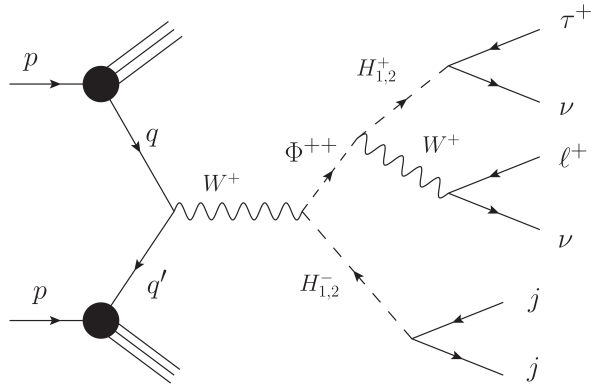}
\caption{The Feynman diagram for the signal process in Scenario-I, where $q$ and $q^\prime$ are partons.}
\label{fig:Signal_ScenarioI}
\end{center}
\end{figure} 

In this scenario, the singly charged scalars decay into $cs$ or $\tau \nu$ dominantly. 
(See Figs.~\ref{fig:Decay_H1} and \ref{fig:Decay_H2}.)
We investigate the process 
$p p \to W^{+\ast} \to \Phi^{++} H_{1,2}^- \to \tau^+ \ell^+ \nu \nu j j$ 
($\ell = e, \mu$). 
The Feynman diagram for the process is shown in Fig.~\ref{fig:Signal_ScenarioI}. 
In this process, the doubly charged scalar $\Phi^{++}$ and one of the singly charged scalars $H_{1,2}^-$ are generated via s-channel $W^{+\ast}$. 
The produced singly charged scalar decays into a pair of jets, 
and $\Phi^{++}$ decays into $\tau^+\ell^+ \nu \nu$ 
through the on-shell pair of the singly charged scalar and $W^+$. 
Thus, in the distribution of the transverse mass
 of $\tau^+ \ell^+ \cancel{E}_T$, 
where $E_T$ is the missing transverse energy, 
we can see the Jacobian peak whose endpoint corresponds to $m_{\Phi}$~\cite{Aoki:2011yk}\footnote{
In general, the transverse mass $M_T$ of $n$ particles is defined as follows. 
\bal
\label{eq:transverse_sum}
& M_T^2 = (E_{T1} + E_{T2} + \cdots + E_{Tn} )^2 
	+ | {\bm p}_{T1} + {\bm p}_{T2} + \cdots + {\bm p}_{Tn} |^2, \\
& E_{Ti}^2 = |{\bm p}_{Ti}|^2 + m_i^2 \quad (i = 1, 2, \cdots, n), 
\eal
where ${\bm p}_{Ti}$ and $m_i$ are 
the transverse momentum and the mass of $i$-th particle, respectively. 
}.
In the present process, furthermore, 
in the distribution of the transverse mass of two jets, 
we can basically see twin Jacobian peaks at $m_{H_1}$ and $m_{H_2}$ ~\cite{Aoki:2011yk}.
Therefore, by using the distributions of $M_T(\tau^+ \ell^+ \cancel{E}_T)$ 
and $M_T(jj)$, 
we can obtain the information on masses of all the charged scalars $H_1^\pm$, $H_2^\pm$, and $\Phi^{\pm\pm}$. 
This is the characteristic feature of the process in this model. 
When we consider the decay of the tau lepton, 
the transverse mass of the decay products of the tau lepton and $\ell^+ \nu \nu$ can be used instead of $M_T(\tau^+ \ell^+ \nu \nu)$. 

In the following, we discuss the kinematics of the process at HL-LHC 
with the numerical evaluation.   
For input parameters, we take the following benchmark values for Scenario-I;  
\beq
\label{eq:Benchmark_value_ScenarioI}
m_{\Phi} = 200~\mathrm{GeV}, \quad
m_{H_1} = 100~\mathrm{GeV}, \quad
m_{H_2} = 120~\mathrm{GeV}, \quad
\tan \beta = 10, \quad
\chi = \frac{ \pi }{ 4 }. 
\eeq
From the LEP data~\cite{Abbiendi:2013hk}, 
the singly charged scalars are heavier than the lower bound of the mass ($80~\mathrm{GeV}$). 
In addition, we take the large $\tan \beta$(=10), so that they satisfy the constraints from flavor experiments~\cite{Enomoto:2015wbn, Haller:2018nnx} 
and LHC Run-I~\cite{Arbey:2017gmh, Aiko:2020ksl}. 

The final state include the tau lepton,  
and we consider the case that the tau lepton decays into $\pi^+ \overline{\nu}$. 
In this case, 
$\pi^+$ flies in the almost same direction of $\tau^+$ in the Center-of-Mass (CM)  frame 
because of the conservation of the angular momentum~\cite{ref:Associated_production}. 
The branching ratio for $\tau^+ \to \pi^+ \overline{\nu}$ is about $11~\%$~\cite{Zyla:2020zbs}, 
and we assume that the efficiency of tagging the hadronic decay of tau lepton is $60~\%$~\cite{Sirunyan:2018pgf}. 
Under the above setup, 
we carry out the numerical evaluation of the signal events 
by using M{\small AD}G{\small RAPH}5\_{\small A}MC@NLO~\cite{Alwall:2014hca}, 
FeynRules~\cite{FeynRules}, and TauDecay~\cite{Hagiwara:2012vz}. 
As a result, about $600$ signal events are expected to be produced at HL-LHC. 
The distributions of the signal events for 
$M_T(\pi^+ \ell^+ \cancel{E}_T)$ and $M_T(jj)$ 
are shown in red line in the left figure of Fig.~\ref{fig:MT_basic_ScenarioI} and 
in the right one, respectively. 
\begin{figure}[h]
\begin{tabular}{c}
	\begin{minipage}[t]{0.5\hsize}
		\begin{center}
		\includegraphics[width=80mm]{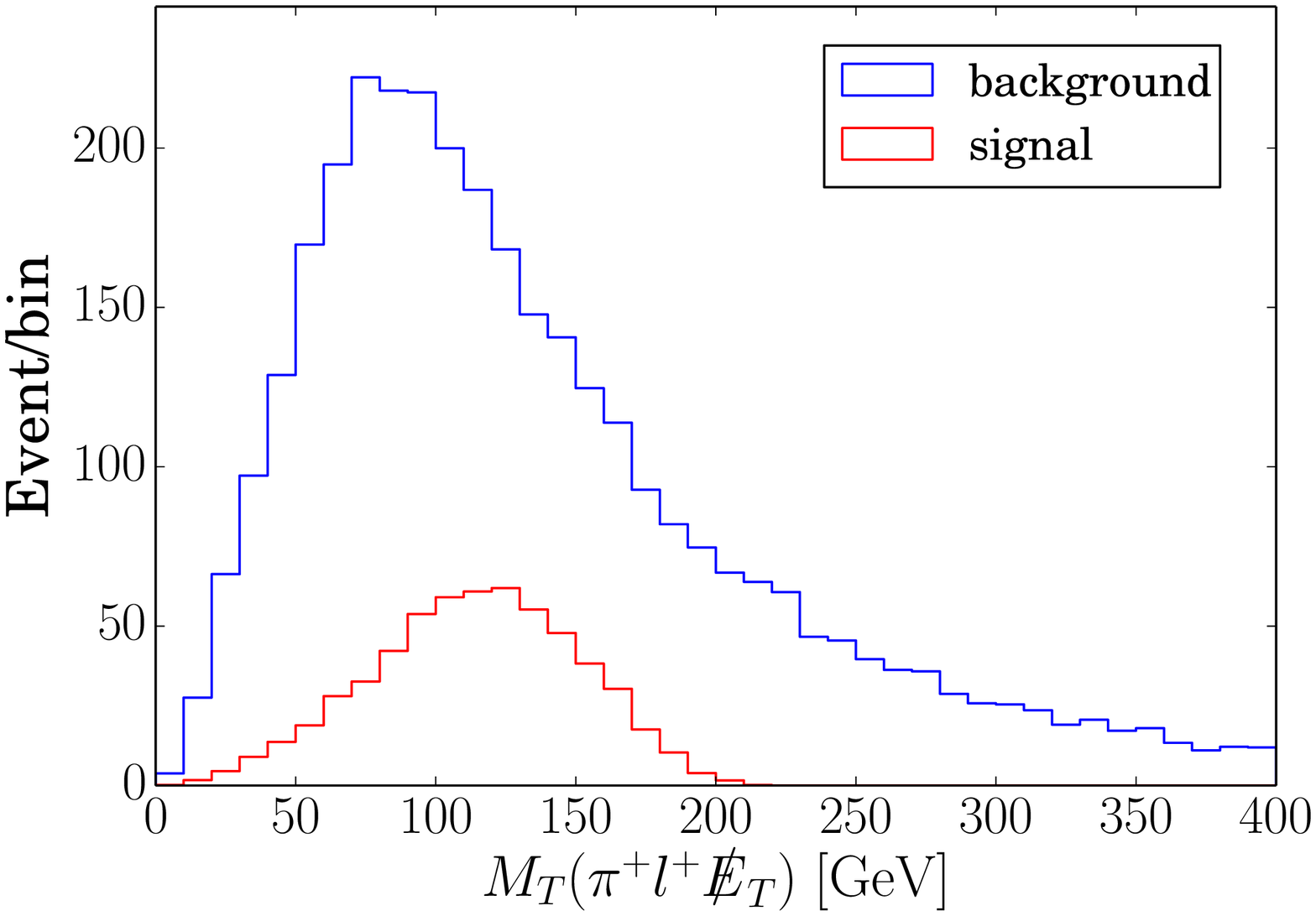}
		\label{fig:MT_basic_ScenarioI}
		\end{center}
	\end{minipage}
	\hspace{10pt}
	\begin{minipage}[t]{0.5\hsize}
		\begin{center}
		\includegraphics[width=80mm]{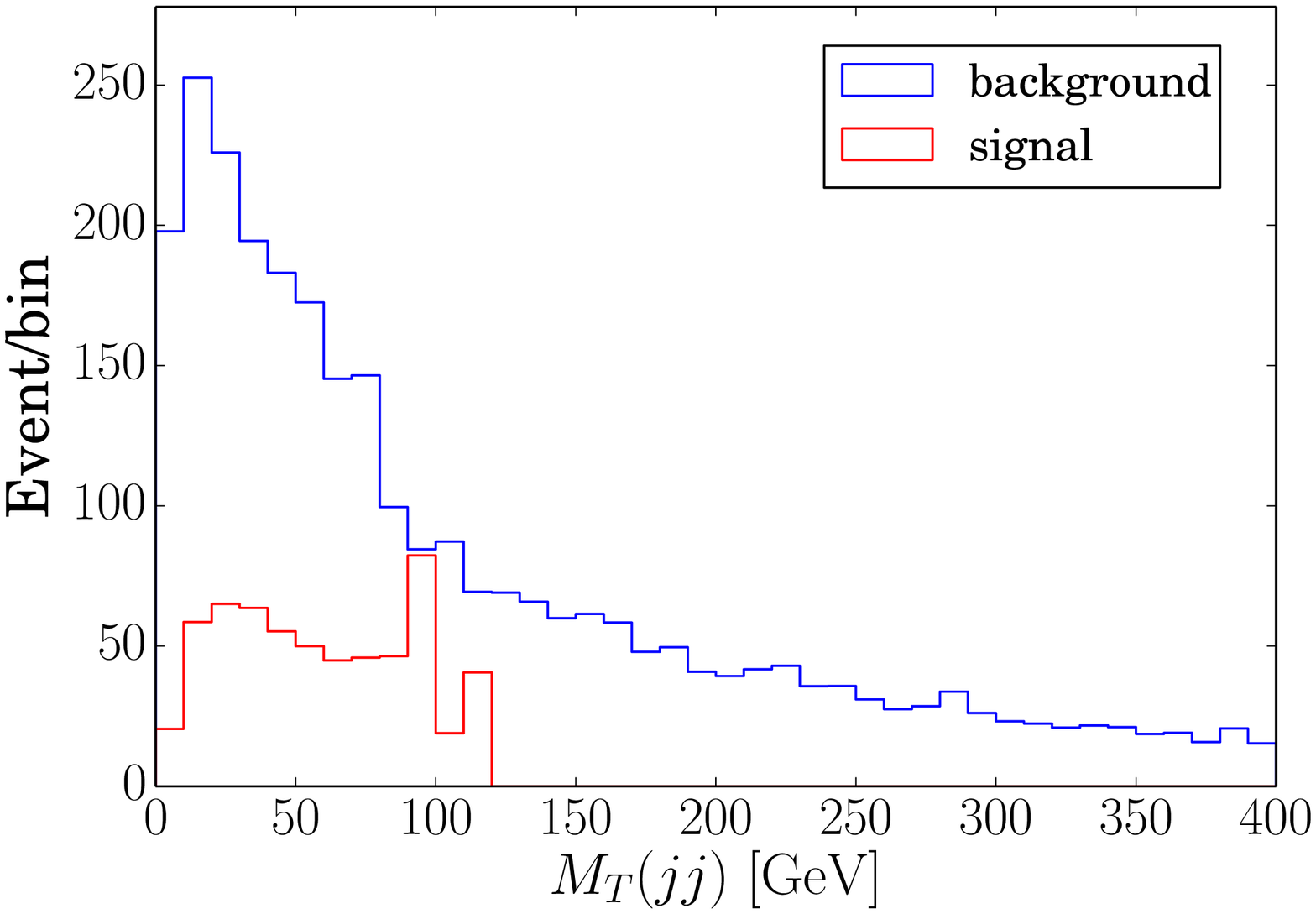}
		\label{fig:MTjj_basic_ScenarioI}
		\end{center}
	\end{minipage}
\end{tabular}
\caption{The distribution of the signal and background events for 
		$M_T(\pi^+ \ell^+ \cancel{E}_T)$ (the left figure) 
		and $M_T(jj)$ (the right one)
		We use the basic cut in Eq.~(\ref{eq:basic_cut}). 
		The width of the bin in the figures 
		is $10~\mathrm{GeV}$. 
		We use the benchmark values in 
		Eq.~(\ref{eq:Benchmark_value_ScenarioI}).}
\label{fig:MT_basic_ScenarioI}
\end{figure}

Next, we discuss the background events and their reduction. 
The main background process is $pp \to W^+ W^+ j j \to \tau^+ \ell^+ \nu \overline{\nu} jj$. 
The leading order of this background process is $\mathcal{O}( \alpha^6 )$ and $\mathcal{O}(\alpha^4 \alpha_s^2)$. 
 For $\mathcal{O}(\alpha^6)$, the vector boson fusion (VBF) and tri-boson production $pp\to W^+ W^+ W^- \to W^+ W^+ j j$ are important. 
On the other hand, for $\mathcal{O}(\alpha^4 \alpha_s^2)$, 
 the main process is t-channel gluon mediated $pp \to q^\ast q^{\prime \ast} \to W^+ W^+ j j$, where $q$ and $q^\prime$ are quarks in internal lines.   
The number of the total background events under the basic cuts in Eq.~(\ref{eq:basic_cut}) is shown in Table~\ref{table:events_scenarioI}. 
Transverse mass distributions of background events for $M_T(\pi^+ \ell^+ \cancel{E}_T)$ and $M_T(jj)$ are shown in the blue line 
in the left figure of Fig.~\ref{fig:MT_basic_ScenarioI} 
and in the right one, respectively.  
The number of the background events is larger than that of the signal.  
Clearly, background reduction has to be performed by additional kinematical cuts. 

\begin{figure}[h]
\begin{tabular}{c}
	\begin{minipage}[t]{0.5\hsize}
		\begin{center}
		\includegraphics[width=80mm]{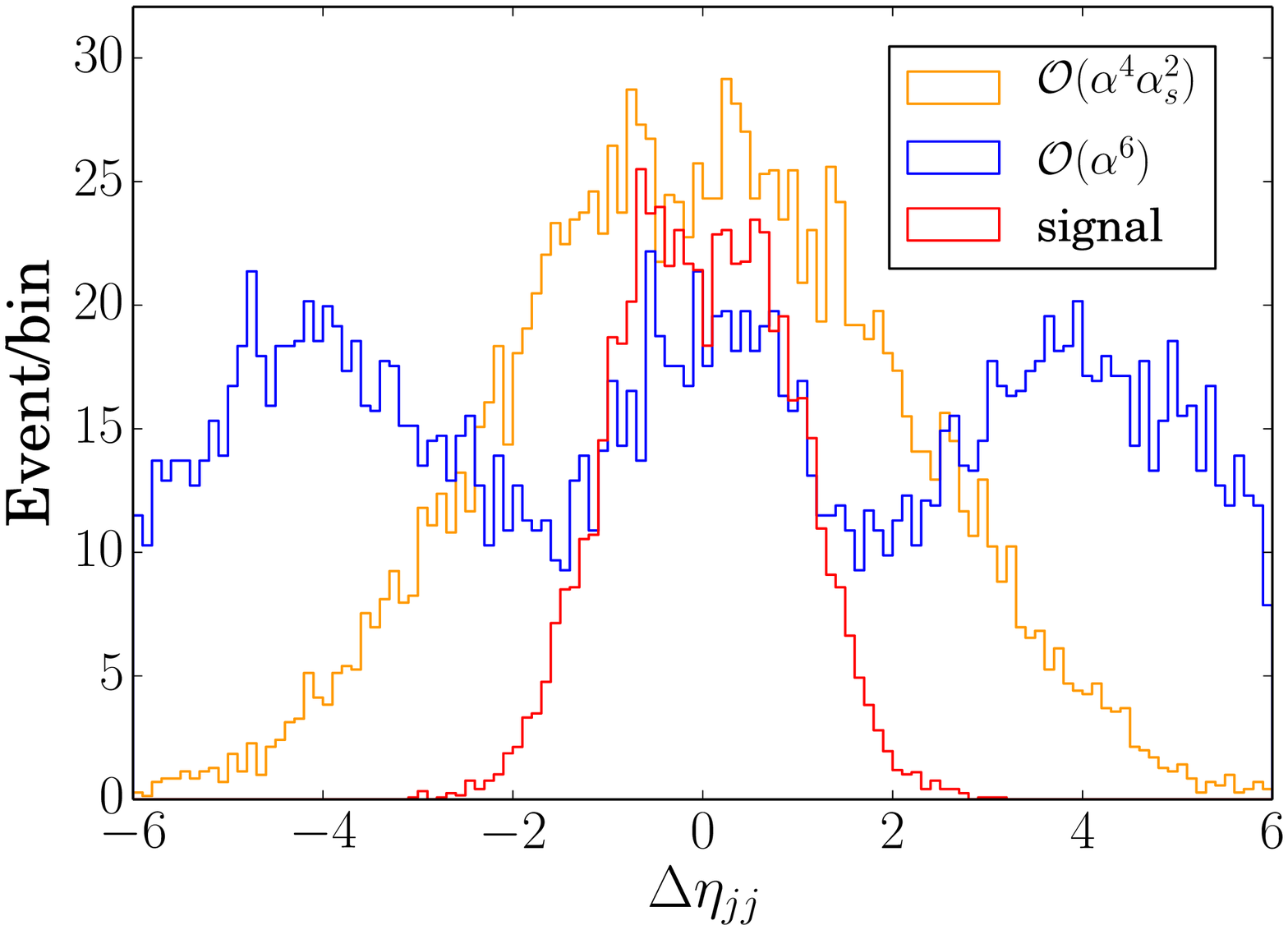}
		\end{center}
	\end{minipage}
	\hspace{10pt}
	\begin{minipage}[t]{0.5\hsize}
		\begin{center}
		\includegraphics[width=80mm]{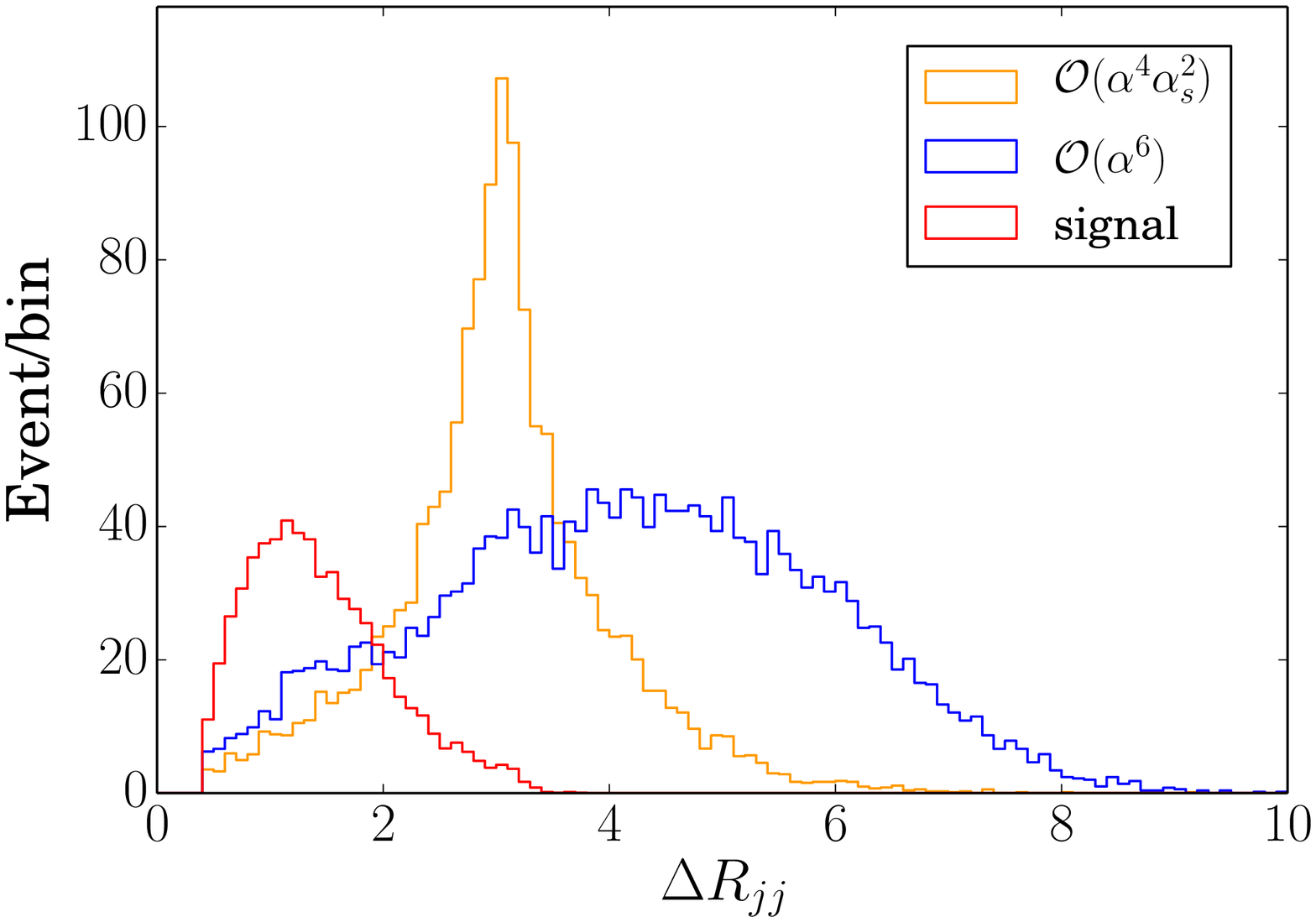}
		\end{center}
	\end{minipage}
\end{tabular}
\begin{center}
\includegraphics[width=80mm]{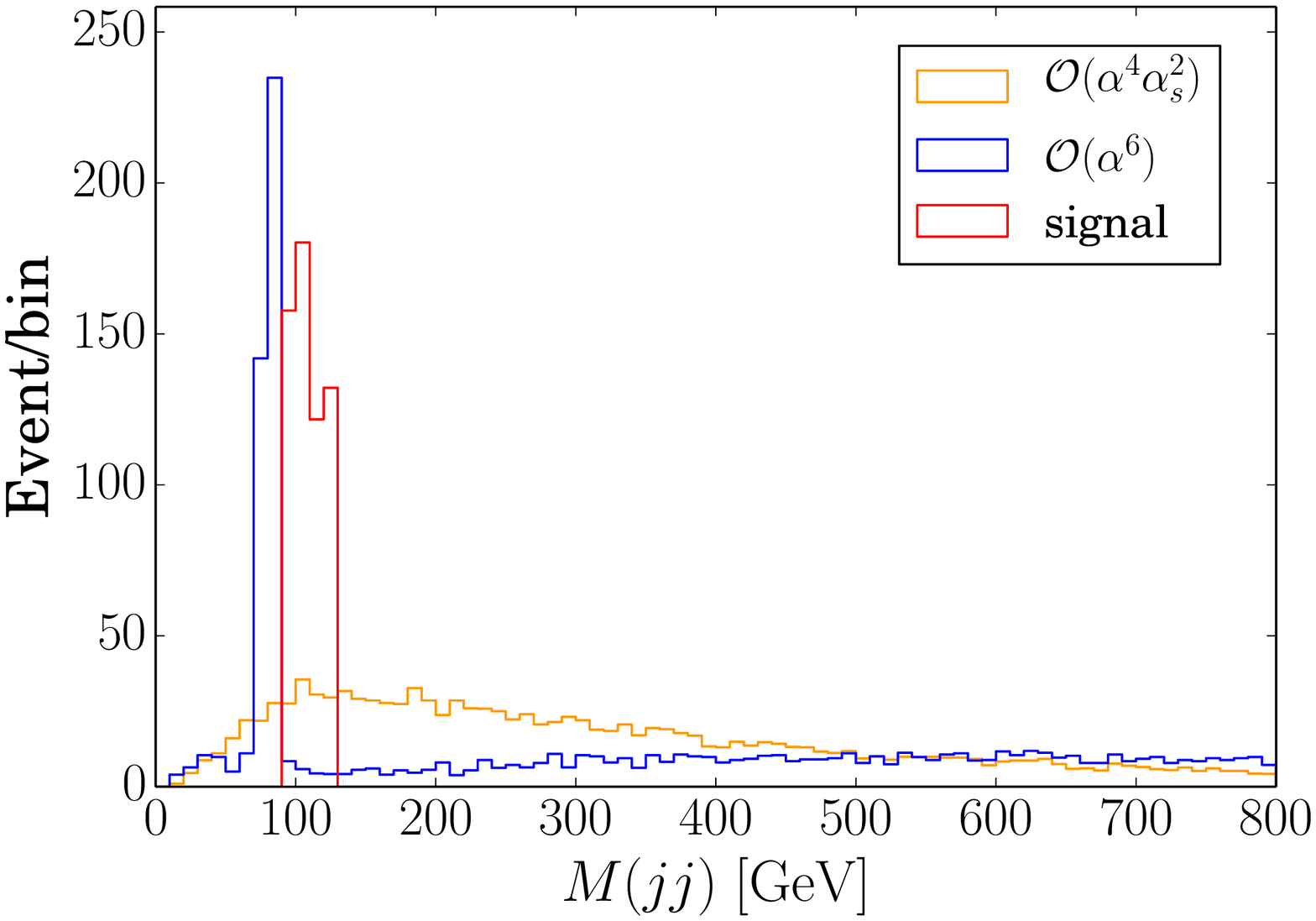}
\end{center}
\caption{The distributions of signal and background events 
for $\Delta \eta_{jj}$ (the upper left figure), 
$\Delta R_{jj}$ (the upper right one), 
and $M_{jj}$ (the bottom one). 
The red lines are those for the signal events. 
The blue (yellow) lines are those for the background events 
of $\mathcal{O}(\alpha^6)$ ($\mathcal{O}(\alpha^4 \alpha_s^2)$). 
In the figures for $\Delta \eta_{jj}$ and $\Delta R_{jj}$, 
we take the width of bins as $0.1$. 
In the figure for $M_{jj}$, the width of bins is $10~\mathrm{GeV}$. 
We use the benchmark values in 
		Eq.~(\ref{eq:Benchmark_value_ScenarioI}).}
\label{fig:some_distributions_ScenarioI}
\end{figure}

First, we impose the pseudo-rapidity cut for a pair of two jets ($\Delta \eta_{jj}$). 
The $\Delta \eta_{jj}$ distributions of the signal and background processes are shown in the upper left figure in Fig.~\ref{fig:some_distributions_ScenarioI}.  
For the signal events, 
the distribution has a maximal value at $\Delta \eta_{jj} = 0$ 
as they are generated via the decay of $H_1^-$ or $H_2^-$. 
On the other hand, for the VBF background, 
two jets fly in the almost opposite directions, 
and each jet flies almost along the beam axis. 
Large $|\Delta \eta_{jj}|$ is then expected to appear~\cite{Ballestrero:2018anz}, 
so that we can use $|\Delta \eta_{jj}| < 2.5$ to reduce the VBF background. 
We note that this kinematical cut is not so effective to reduce other $\mathcal{O}(\alpha^6)$ and $\mathcal{O}(\alpha^4 \alpha_s^2)$ processes 
because in these background, the distribution are maximal at $\Delta \eta_{jj}=0$. 

Second, we impose the angular distance cut for a pair of two jets ($\Delta R_{jj}$). 
The $\Delta R_{jj}$ distributions of the signal and background processes are shown in the upper right figure in Fig.~\ref{fig:some_distributions_ScenarioI}. 
For the signal events, 
the distribution has a maximal value at $\Delta R_{jj} \simeq 1.0$. 
On the other hand, 
for the $\mathcal{O}(\alpha^4 \alpha_s^2)$ background events, 
$\Delta R_{jj}$ has a peak at $\Delta R_{jj} \sim \pi$. 
In addition, in the $\mathcal{O}(\alpha^6)$ ones, 
$\Delta R_{jj}$ has large values between $3$ and $6$. 
Therefore, for $\Delta R_{jj} < 2$, 
the background events are largely reduced 
while the almost all signal events remains. 

Third, we impose invariant mass cut for a pair of two jets ($M_{jj}$). 
The $M_{jj}$ distributions of the signal and background processes are shown 
in the bottom figure in Fig.~\ref{fig:some_distributions_ScenarioI}. 
For the signal events, as they are generated via the decay of the singly charged scalars, 
the distribution has twin peaks at the masses of $H_1^\pm$ and $H_2^\pm$  ($100~\mathrm{GeV}$ and $120~\mathrm{GeV}$). 
On the other hand, for the background events, 
the jets are generated via on-shell $W$ or t-channel diagrams. 
Then, the distribution of the background has a peak at the $W$ boson mass ($\sim 80~\mathrm{GeV}$).
Thus, the kinematical cut $90~\mathrm{GeV} < M_{jj} < 180~\mathrm{GeV}$ is so effective to reduce the background events.  
We note that this reduction can only be possible when we already know 
some information on the masses of the singly charged scalars. 

We summarize three kinematical cuts for the background reduction. 
\bal
\label{eq:kinematical_cuts}
&(\mathrm{i}) \quad |\Delta \eta_{jj}| < 2.5, \\
&( \mathrm{ii}) \quad \Delta R_{jj} < 2, \\
&(\mathrm{iii}) \quad 90~\mathrm{GeV} < M_{jj} < 180~\mathrm{GeV}, 
\eal

\begin{table}[h]
\begin{center}
\begin{tabular}{c|c|c|c|}
 & signal $S$ & background $B$ & $S/\sqrt{S+B}$ \\ \hline
\begin{tabular}{c}
Basic cuts \\ 
(Eq.~(\ref{eq:basic_cut})) \\
\end{tabular}
	 & 592 & 3488 & 9.3 \\ \hline
\begin{tabular}{c}
 Basic cuts (Eq.~(\ref{eq:basic_cut}))\\
and $\Delta R_{jj} < 2$, $|\Delta \eta_{jj}| < 2.5$ \\
\end{tabular}
	& 487 & 412 & 16 \\ \hline
\begin{tabular}{c}
All cuts \\
( Eq.~(\ref{eq:basic_cut}) and Eq.~(\ref{eq:kinematical_cuts}) )\\
\end{tabular}
	& 487 & 75 & 20 \\ \hline
\end{tabular}
\caption{Numbers of signal event and background events at HL-LHC in Scenario I. 
In the first column, the number of events under only the basic cuts are shown. 
The number of events under the all cuts are shown in the second column. 
We use the benchmark values in 
		Eq.~(\ref{eq:Benchmark_value_ScenarioI}).}
\label{table:events_scenarioI}
\end{center}
\end{table}

\begin{figure}[h]
\begin{tabular}{c}
	\begin{minipage}[t]{0.5\hsize}
		\begin{center}
		\includegraphics[width=80mm]{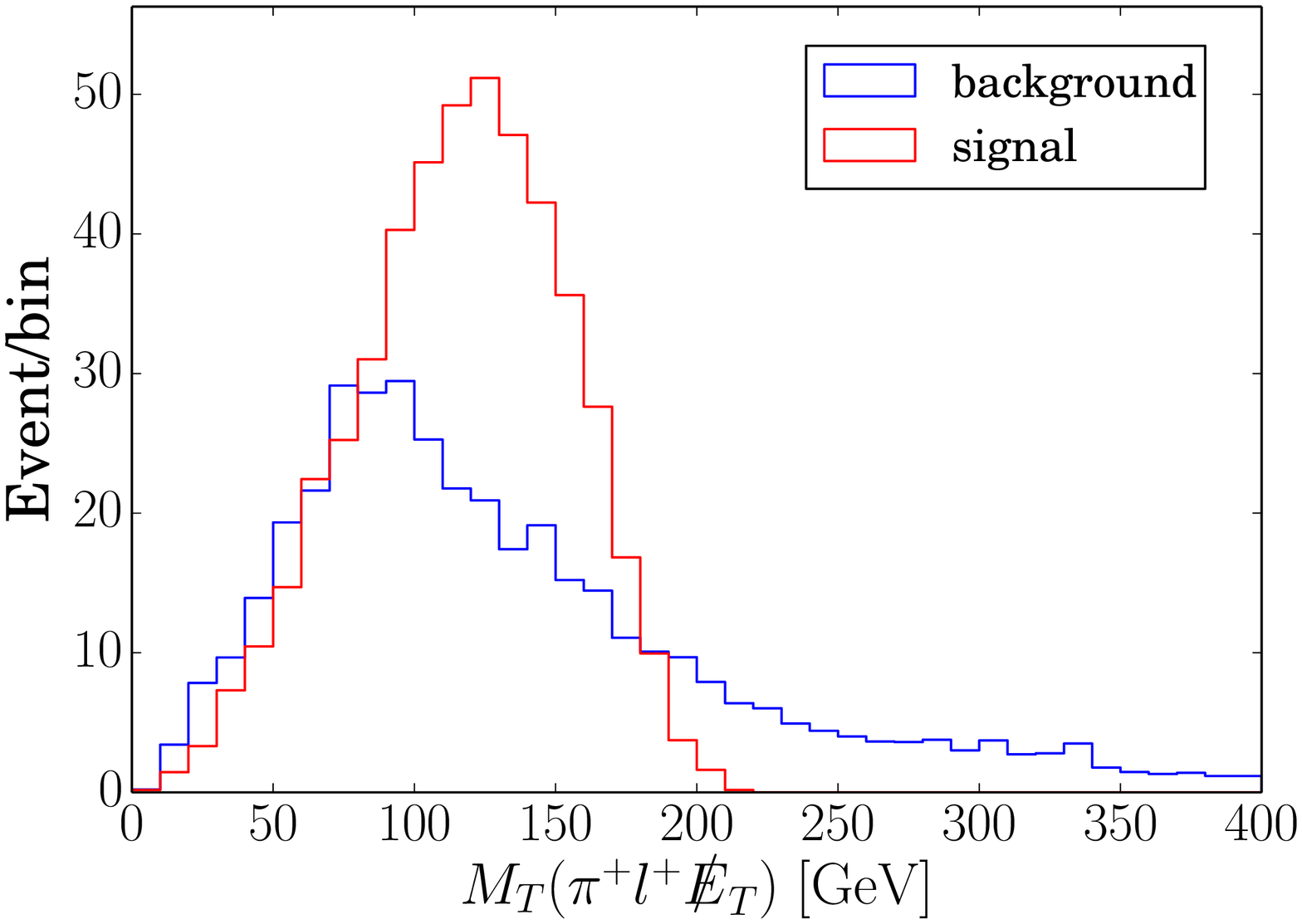}
		\end{center}
	\end{minipage}
	\begin{minipage}[t]{0.5\hsize}
		\begin{center}
		\includegraphics[width=80mm]{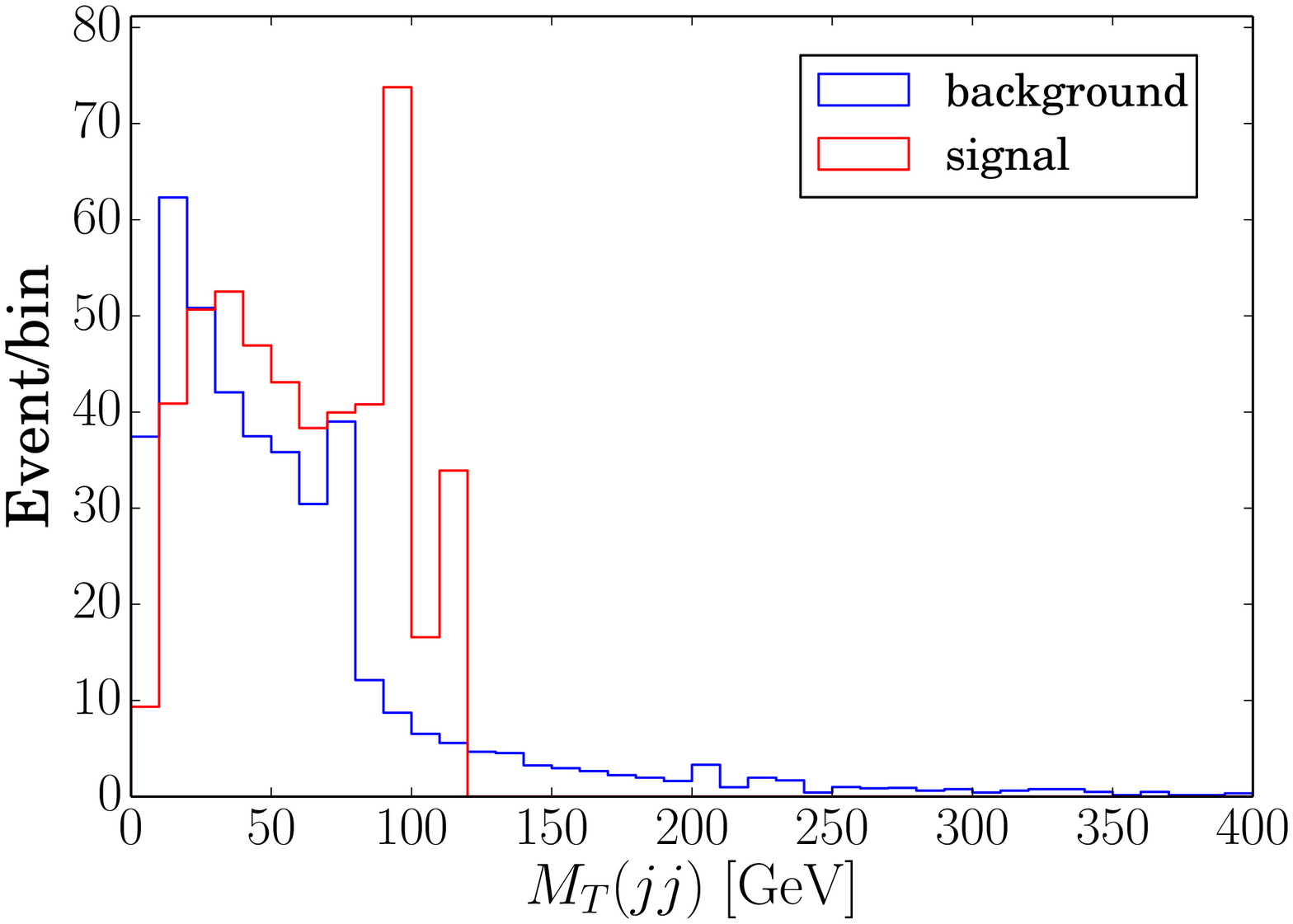}
		\end{center}
	\end{minipage}
\end{tabular}
\caption{The distribution of the signal and background events for 
		$M_T(\pi^+ \ell^+ \cancel{E}_T)$ (the left figure) 
		and $M_T(jj)$ (the right one)
		We use the basic cuts in Eq.~(\ref{eq:basic_cut}), 
		$|\Delta \eta_{jj}| < 2.5$, 
		and $\Delta R_{jj} < 2$. 
		The width of bins in the figures 
		is $10~\mathrm{GeV}$.
		We use the benchmark values in 
		Eq.~(\ref{eq:Benchmark_value_ScenarioI}).}
\label{fig:MT_caseI_ScenarioI}
\end{figure}

Let us discuss how the backgrounds can be reduced by using the first two kinematical cuts (i) and (ii), 
in addition to the basic cuts given in Eq.~(\ref{eq:basic_cut}). 
This corresponds to the case that we do not use the information on the masses of the singly charged scalars. 
The results are shown in the third column of Table~\ref{table:events_scenarioI}. 
In this case, about $88~\%$ of the background events are reduced, 
while about $82~\%$ of the signal events remain. 
We obtain the significance as $S/\sqrt{S+B} =16$. 
The distributions for $M_T ( \pi^+ \ell^+ \cancel{E}_T)$ and $M_T(jj)$ 
are shown in Fig.~\ref{fig:MT_caseI_ScenarioI}. 
In the left figure of Fig.~\ref{fig:MT_caseI_ScenarioI}, 
we can see the Jacobian peak of 
$M_T ( \pi^+ \ell^+ \cancel{E}_T)$. 
Consequently, the signal process can be detected at HL-LHC in Scenario-I of Eq.~(\ref{eq:Benchmark_value_ScenarioI}). 
However, the endpoint of the signal is unclear due to the background events, 
so that it would be difficult to precisely decide the mass of $\Phi^{++}$. 
On the other hand, we can see the twin Jacobian peaks of 
$M_T ( jj )$ in the right figure of Fig.~\ref{fig:MT_caseI_ScenarioI}. 
Therefore, we can also obtain information on masses of both the singly charged scalars. 
In this way, all the charged scalar states $\Phi^{\pm\pm}$, $H_1^\pm$, and $H_2^\pm$ can be detected and their masses may be obtained to some extent.

\begin{figure}[h]
\begin{tabular}{c}
	\begin{minipage}[t]{0.5\hsize}
		\begin{center}
		\includegraphics[width=80mm]{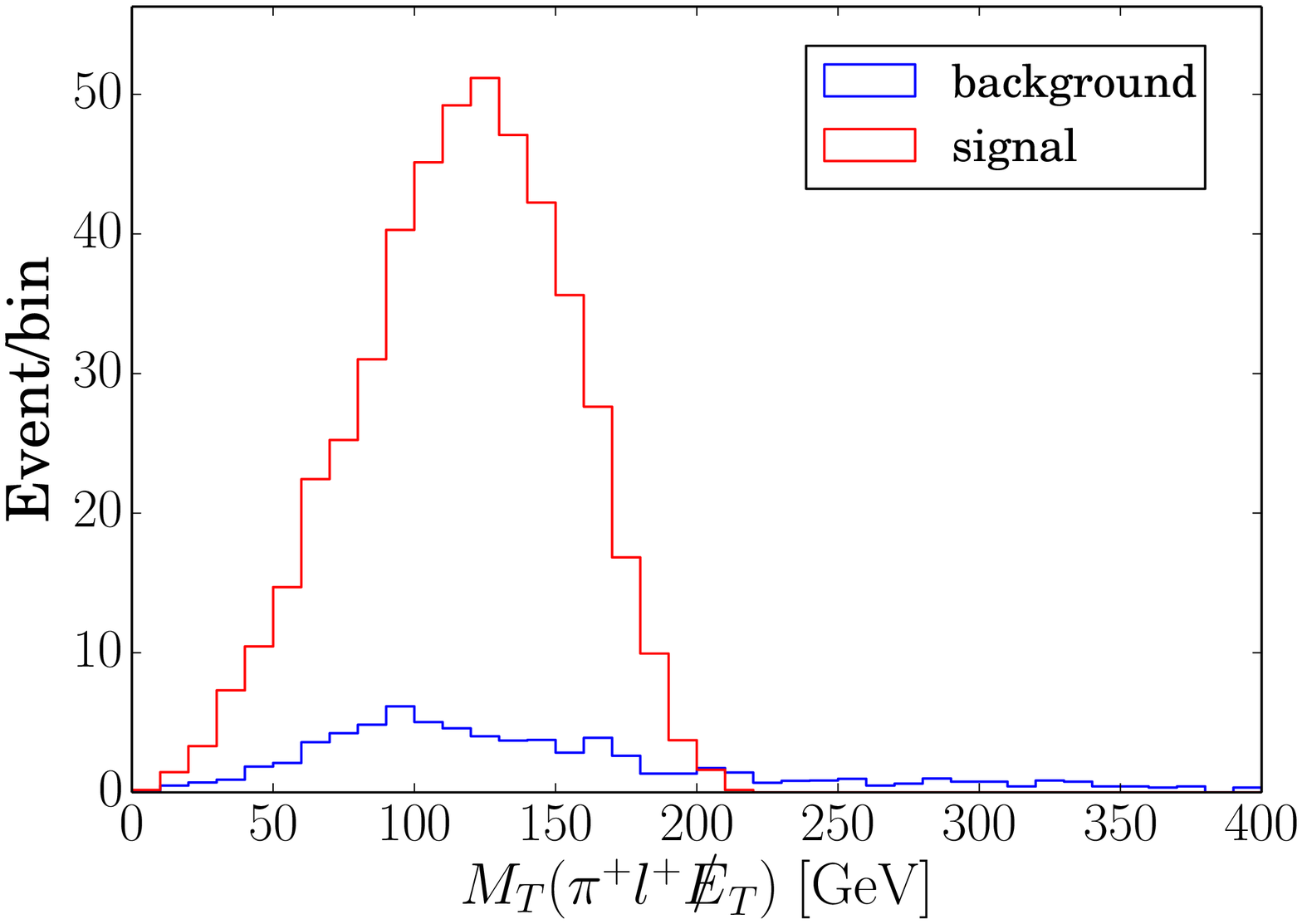}
		\end{center}
	\end{minipage}
	\begin{minipage}[t]{0.5\hsize}
		\begin{center}
		\includegraphics[width=80mm]{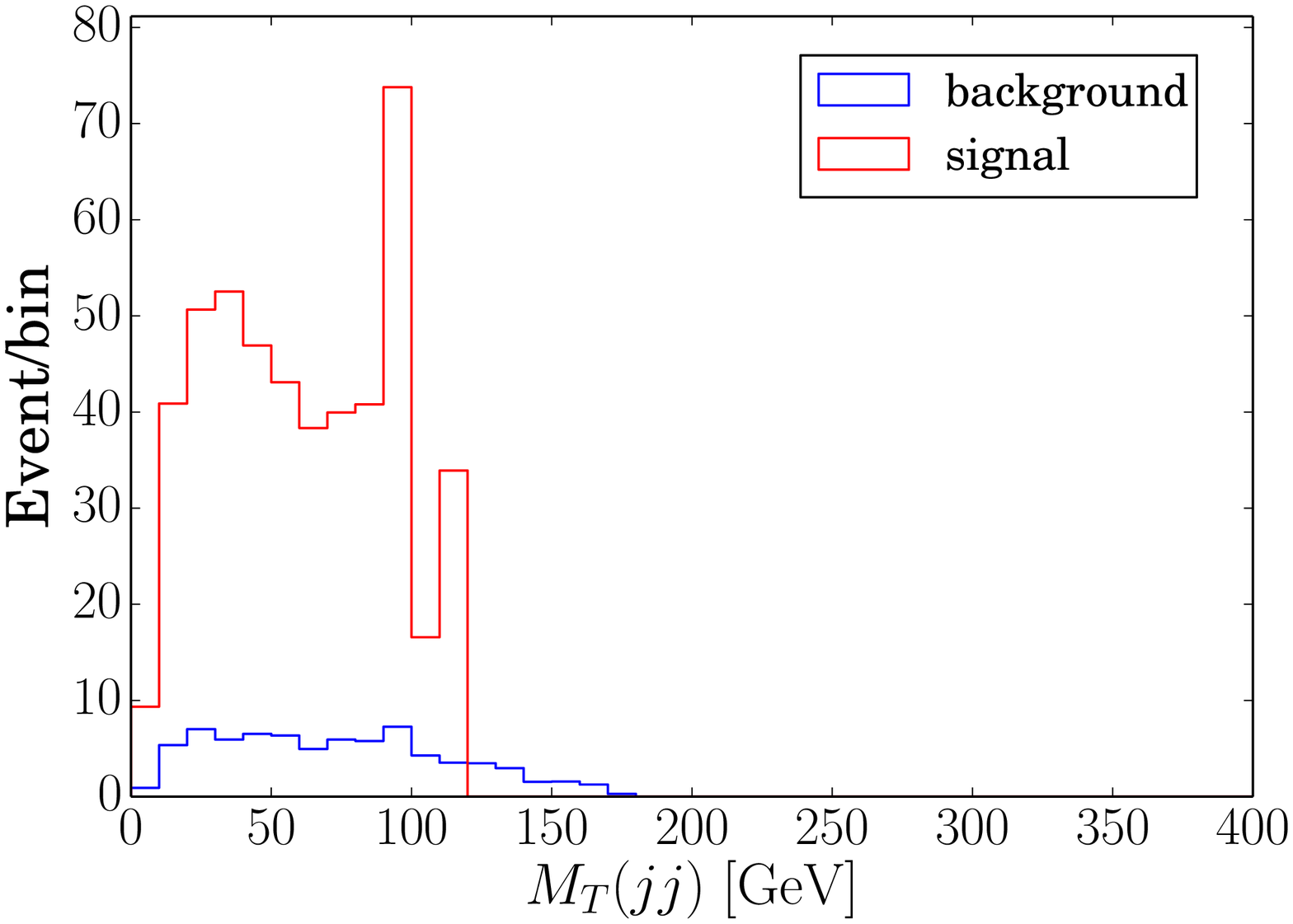}
	\end{center}
	\end{minipage}
\end{tabular}
\caption{The distribution of the signal and background events for 
		$M_T(\pi^+ \ell^+ \cancel{E}_T)$ (the left figure) 
		and $M_T(jj)$ (the right figure)
		We use the basic cut in Eq.~(\ref{eq:basic_cut}) 
		and all the kinematical cuts 
		in Eq.~(\ref{eq:kinematical_cuts}). 
		The width of the bin in the figures 
		is $10~\mathrm{GeV}$.}
\label{fig:MT_caseII_ScenarioI}
\end{figure}

Furthermore, if we impose all the kinematical cuts (i), (ii), and (iii) with the basic cuts, 
the backgrounds can be further reduced. 
The results are shown in the fourth column of Table~\ref{table:events_scenarioI}. 
The number of signal events are same with that in the previous case. 
On the other hand, the background reduction is improved, 
and $98~\%$ of the background events are reduced. 
The significance is also improved as $S/\sqrt{S+B} = 20$. 
Distributions for $M_T ( \pi^+ \ell^+ \cancel{E}_T)$ and $M_T(jj)$ 
are shown in Fig~\ref{fig:MT_caseII_ScenarioI}. 
In the left figure of Fig~\ref{fig:MT_caseII_ScenarioI}, 
we can see that there are only few background events 
around the end point of Jacobian peak $M_T ( \pi^+ \ell^+ \cancel{E}_T)$. 
Thus, it would be expected we obtain the more clear information on $m_{\Phi}$ than that from the case where only (i) and (ii) are imposed as additional kinematical cuts. 
We can also clearly see the twin Jacobian peaks 
in the right figure of Fig~\ref{fig:MT_caseII_ScenarioI}, 
and a large improvement can be achieved for the determination of the masses of both the singly charged scalar states. 

Before closing Subsection A, 
we give a comment about the detector resolution. 
In the process, 
the transverse momenta of jets ($p_T^j$) are mainly distributed 
between $0$ and $200~\mathrm{GeV}$, 
and the typical value of them is about $100~\mathrm{GeV}$. 
According to Ref.~\cite{Aad:2020flx}, 
at the current ATLAS detector, 
the energy resolution for $p_T^j \simeq 100~\mathrm{GeV}$ 
is about $10~\%$. 
In Figs.~\ref{fig:MT_basic_ScenarioI}-\ref{fig:MT_caseII_ScenarioI}, 
we take the width of bins as $10~\mathrm{GeV}$. 
Therefore, 
it would be possible that 
the twin Jacobian peaks in the distribution for $M_T (jj)$ 
overlap each other and 
they looks like one Jacobian peak with the unclear endpoint at the ATLAS detector if the mass differences is not large enough. 
Then, it would be difficult to obtain the information on 
both $m_{H_1}^{}$ and $m_{H_2}^{}$ 
from the transverse momentum distribution.  
Even in this case, 
it would be able to obtain the hint for the masses by investigating the process. 
In our analysis, 
we did not consider the background where the $Z$ boson decays into dijet such as $q q \to Z^\ast \to Z h \to j j \tau \overline{\tau}
\to j j \pi^+ \overline{\nu}_\tau \ell^- \nu_{\tau} \overline{\nu}_\ell$, 
which can be expected to be reduced by veto the events of $M_{jj}$ at the $Z$ boson mass and the cut of the transverse mass $M_T (\pi^+ \ell^+ \cancel{E}_T)$ below $125~\mathrm{GeV}$. 
It does not affect the Jacobian peak and the endpoint at the mass of doubly charged scalar boson $\Phi^{\pm\pm}$. 

\subsection{Scenario-II}

\begin{figure}[h]
\begin{center}
\includegraphics[width=110mm]{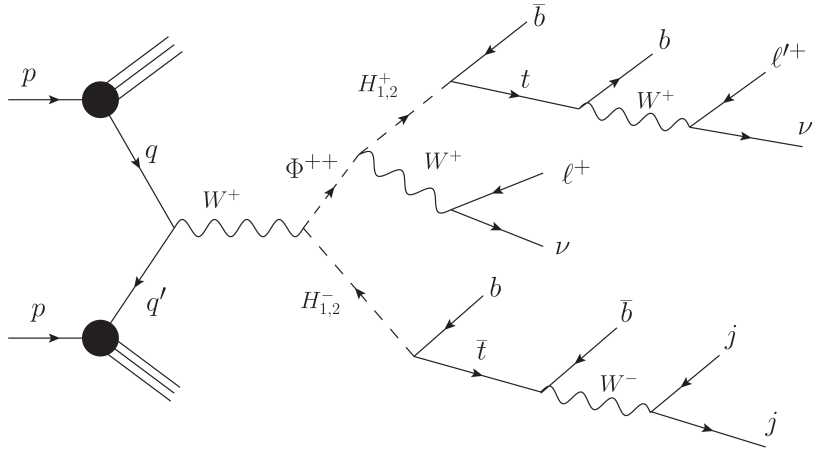}
\caption{The Feynman diagram for the signal process in Scenario-II, where $q$ and $q^\prime$ are partons.}
\label{fig:Signal_ScenarioII}
\end{center}
\end{figure}

In this scenario, the singly charged scalars predominantly decay into $tb$ with the branching ratio almost $100~\%$. 
We investigate the signal 
$pp \to W^{+\ast} \to \Phi^{++} H_{1,2}^- \to t \overline{t}b \overline{b} \ell^+ \nu 
\to b b \overline{b} \overline{b} \ell^+ \ell^{\prime +} \nu \nu j j$ 
($\ell, \ell^\prime = e, \mu$). 
The Feynman diagram for the process is shown in Fig.~\ref{fig:Signal_ScenarioII}. 
The decay products of $\Phi^{++}$ and $H_{1,2}^\pm$ are 
$b \overline{b} \ell^+ \ell^{\prime +} \nu \nu$ and $b \overline{b} j j$, respectively. 
Therefore, in the same way as Scenario-I, 
we can obtain information on masses of all the charged scalars 
by investigating the transverse distributions of signal and background events for  $M_T(b \overline{b} \ell^+ \ell^{\prime +} \nu \nu)$ and $M_T(b \overline{b} j j)$. 
However, in the Scenario-II, 
decay products of both $\Phi^{++}$ and $H_{1,2}^-$ include a $b \overline{b}$ pair, 
and it is necessary to distinguish the origin of the two $b \overline{b}$ pairs. 
We suggest the following two methods of the distinction. 

In the first method, we use the directions of $b$ and $\overline{b}$. 
In the process, $\Phi^{++}$ and $H_{1,2}^-$ are generated with momenta in the opposite directions, 
and decay products fly along the directions of each source particle. 
The both of two $W$ bosons generated via the decay of $\Phi^{++}$ decay into charged leptons and neutrinos, 
while the $W$ boson via the decay of $H_{1,2}$ decays into a pair of jets. 
By using this topology of the process, we can distinguish the origin of two $b \overline{b}$ pairs. 
The $b \overline{b}$ pair which flies along the charged leptons $\ell^+$ and $\ell^{\prime +}$ 
(and flies along the almost opposite direction of a pair of jets) 
comes from the decay of $\Phi^{++}$. 
The other $b \overline{b}$ pair is the decay product of $H_{1,2}^-$.

In the second method, we use the transverse momenta of $b$ and $\overline{b}$. 
As shown in the Feynman diagram in Fig.~\ref{fig:Signal_ScenarioII}, 
in the decay chain of $\Phi^{++}$, 
$b$ is generated via the decay of the top quark while $\overline{b}$ is generated via the decay of the singly charged scalars from the decay of $\Phi^{++}$. 
On the other hand, in the decay chain of $H_{1,2}^-$, 
$b$ is generated via the decay of the singly charged scalars while $\overline{b}$ is generated via the decay of the anti-top quark. 
Therefore, when the singly charged scalars are heavy enough to satisfy the inequality, 
\beq
\label{eq:Heavier_H_inequality}
m_{H_{1,2}} - m_t - m_b > m_t - m_W - m_b, 
\eeq
the typical value of the transverse momentum of $b$ from $H_{1,2}^-$ is larger than that of $b$ from the top quark. 
In the same way, the typical value of transverse momentum of $\overline{b}$ from $H_{1,2}^+$ is larger than that of $\overline{b}$ from the anti-top quark. 
Therefore, in this case,  
we can construct the $b \overline{b}$ pair which mainly comes from the decay of $\Phi^{++}$ 
by selecting $b$ with the smaller transverse momentum and $\overline{b}$ with the larger transverse momentum. 
The other $b \overline{b}$ pair comes from the decay of $H_{1,2}^-$. 
On the contrary, when the singly charged scalars are light enough to satisfy the inequality, 
\beq
\label{eq:Lighter_H_inequality}
m_{H_{1,2}} - m_t - m_b < m_t - m_W - m_b, 
\eeq
the typical value of the transverse momentum of $b$ ($\overline{b}$) from $H_{1,2}^-$ ($H_{1,2}^+$) is smaller than that of $b$ ($\overline{b}$) 
from the top quark (the anti-top quark). 
Therefore, in the case where the singly charged scalar is so light that they satisfy the inequality in Eq.~(\ref{eq:Lighter_H_inequality}), 
we can construct the $b \overline{b}$ pair which mainly comes from the decay of $\Phi^{++}$ 
by selecting $b$ with the larger transverse momentum and $\overline{b}$ with the smaller transverse momentum. 
The other $b \overline{b}$ pair comes from the decay of $H_{1,2}^-$. 
Finally, when the masses of singly charged scalars are around $250~\mathrm{GeV}$, 
they satisfy the equation, 
\beq
m_{H_{1,2}} - m_t - m_b \simeq m_t - m_W - m_b. 
\eeq
Then, the typical values of the transverse momenta of two $b$ are similar, and those of two $\overline{b}$ are also similar. 
Therefore, 
we can construct the correct $b \overline{b}$ pair only partly by using the above method, 
and it is not so effective. 
In this case, the first method explained in the previous paragraph is needed. 

In the following, 
we discuss the signal and the background events at HL-LHC with the numerical calculation. 
In the numerical evaluation, we take the following benchmark values as Scenario-II. 
\beq
\label{eq:benchmark_value_ScenarioII}
m_{\Phi} = 300~\mathrm{GeV}, \quad
m_{H_1} = 200~\mathrm{GeV}, \quad
m_{H_2} = 250~\mathrm{GeV}, \quad
\tan \beta = 3, \quad
\chi = \frac{ \pi }{ 4 }. 
\eeq
For $\tan \beta = 3$, 
the lower bound on the masses of singly charged scalars is about $170~\mathrm{GeV}$ as mentioned in the end of Sec.~II. 
Then, this benchmark values satisfy the experimental constraints on singly charged scalars. 
In addition, we adopt the assumption about the collider performance 
at HL-LHC in Eq.~(\ref{eq:HL-LHC}), 
and we use the basic kinematical cuts in Eq.~(\ref{eq:basic_cut}). 
The final state of the signal includes two bottom quarks and two anti-bottom quarks, 
and we assume that the efficiency of the b-tagging is $70~\%$ per one bottom or anti-bottom quark~\cite{Sirunyan:2017ezt}. 
Thus, the total efficiency of the b-tagging in the signal event is about $24~\%$.  
In the numerical calculation, we use 
M{\small AD}G{\small RAPH}5\_{\small A}MC@NLO~\cite{Alwall:2014hca}, 
FeynRules~\cite{FeynRules}. 

As a result, 
145 events are expected to appear at HL-LHC as shown in Table~\ref{table:events_scenarioII}. 
In this benchmark scenario of Eq.~(\ref{eq:benchmark_value_ScenarioII}), 
$H_1^\pm$ is so light that we can use the distinction of the $b\overline{b}$ pair 
in the case where $m_{H_1} - m_t - m_b < m_t - m_b - m_W$. 
Therefore,
we can construct the $b \overline{b}$ pair which mainly comes from the decay of $H_1^-$ 
by selecting $b$ with the smaller transverse momentum and $\overline{b}$ with the larger transverse momentum. 
On the other hand, 
the mass of $H_2^\pm$ is $250~\mathrm{GeV}$, and it satisfies the equation $m_{H_2} - m_t - m_b \simeq m_t - m_b - m_W$. 
Therefore, the selection of $b$ and $\overline{b}$ by their transverse momenta is partly effective 
in the signal where $H_2^-$ is produced with $\Phi^{++}$ via $W^{+\ast}$.\footnote{
We note that we assume some information on the mass of singly charged scalars to select the kinematical cuts. }  

\begin{figure}[h]
\begin{tabular}{c}
	\begin{minipage}[t]{0.5\hsize}
		\begin{center}
		\includegraphics[width=70mm]{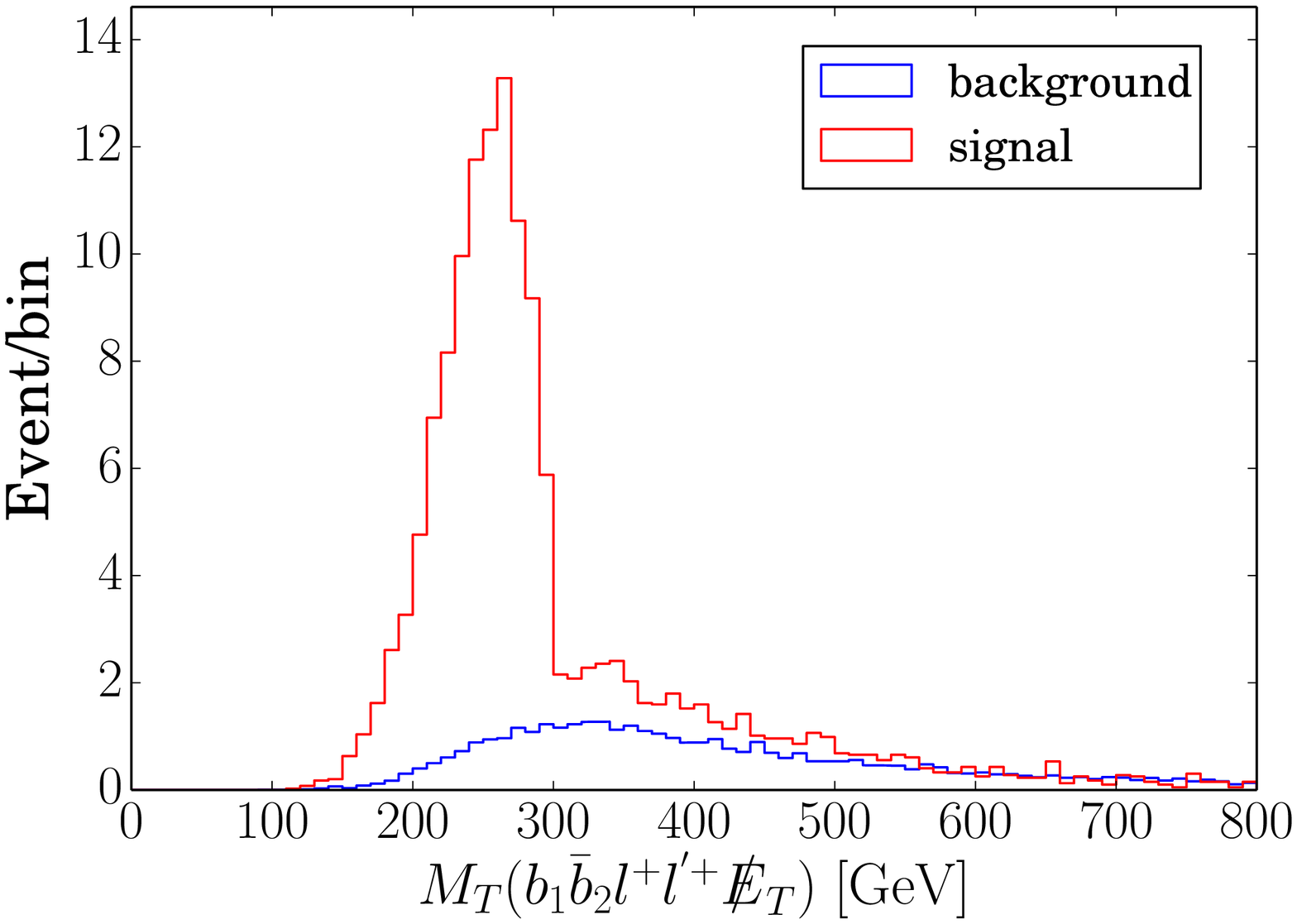}
		\end{center}
	\end{minipage}
	\hspace{10pt}
	\begin{minipage}[t]{0.5\hsize}
		\begin{center}
		\includegraphics[width=70mm]{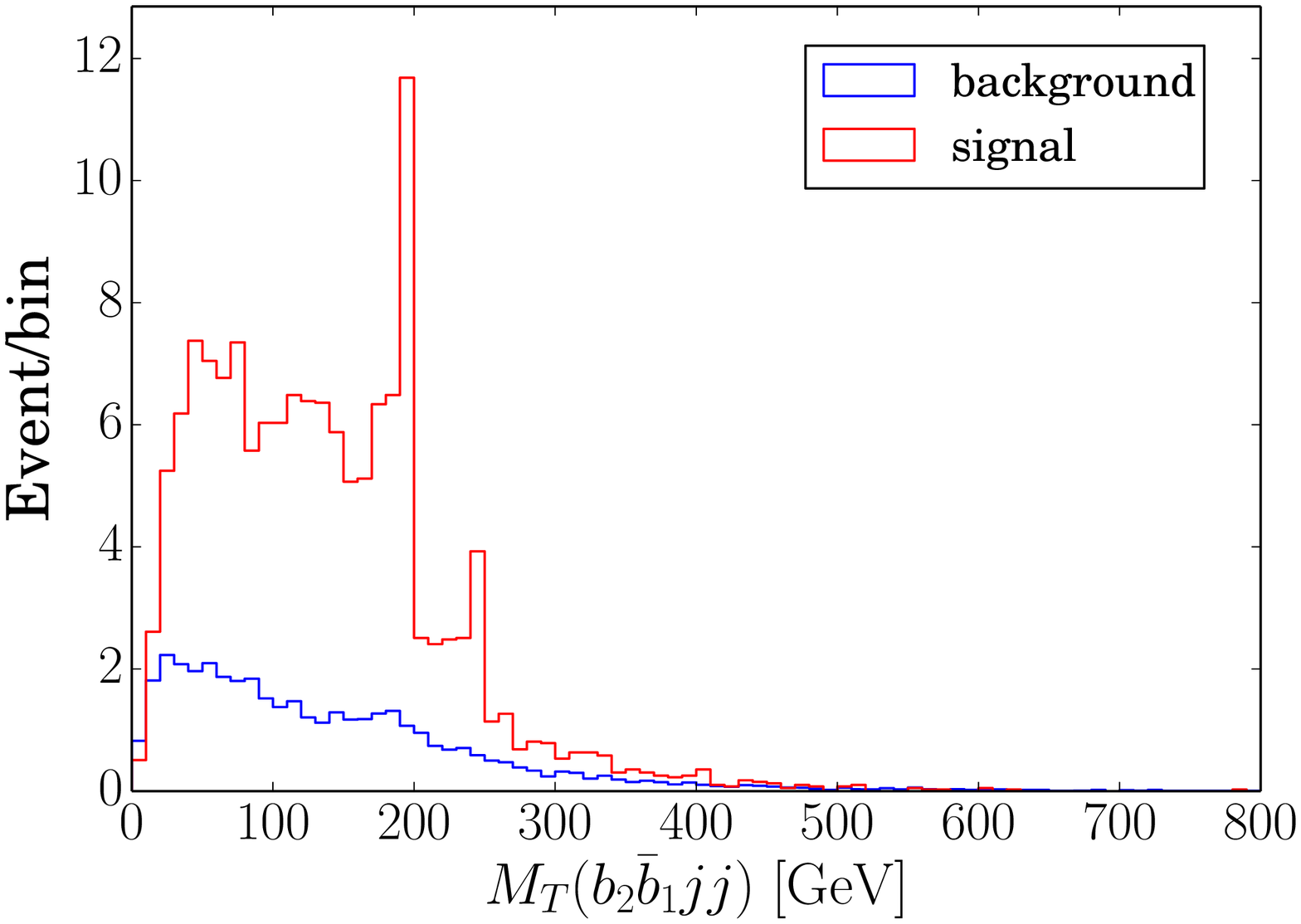}
		\end{center}
	\end{minipage}
\end{tabular}
\caption{The distribution of $M_T(b_1 \overline{b}_2 \ell^+ \ell^{\prime +} \cancel{E}_T)$ 
		(the left one) and $M_T(b_2 \overline{b}_1 jj)$ (the right one)
		in the signal and background events under the kinematical cuts in Eq.~(\ref{eq:basic_cut}). 
		In the figures, the width of bins is $10~\mathrm{GeV}$. 
		We use the benchmark values in 
		Eq.(\ref{eq:benchmark_value_ScenarioII}).}
		\label{fig:signal_plot_ScenarioII}
\end{figure}

In Figs.~\ref{fig:signal_plot_ScenarioII}, 
we show the distributions of 
$M_T(b_1 \overline{b}_2 \ell^+ \ell^{\prime +} \cancel{E}_T)$ 
and $M_T(b_2 \overline{b}_1 j j)$, where 
$b_1$ ($\overline{b}_1$) is the bottom quark (anti-bottom quark) 
with the larger transverse momentum and $b_2$ ($\overline{b}_2$) is the other. 
In the left figure of Fig.~\ref{fig:signal_plot_ScenarioII}, 
the endpoint of the Jacobian peak is not so sharp 
because the selection of the $b \overline{b}$ pairs do not work well 
in the associated production of $\Phi^{++}$ and $H_2^-$. 
In the right figure of Fig.~\ref{fig:signal_plot_ScenarioII}, 
we can see the twin Jacobian peaks at the masses of the singly charged scalars. 
However, the number of events around the Jacobian peaks, especially the one due to $H_2^\pm$, are small, 
and it would be difficult to obtain information on masses 
form the distribution for $M_T(b_2 \overline{b}_1 jj)$. 
In order to obtain the clearer information on $m_{H_{1,2}}$, 
we can use the invariant mass of $b_2 \overline{b}_1 j j$ instead of $M_T(b_2 \overline{b}_1 j j)$. 

\begin{figure}[h]
\begin{center}
\includegraphics[width = 70mm]{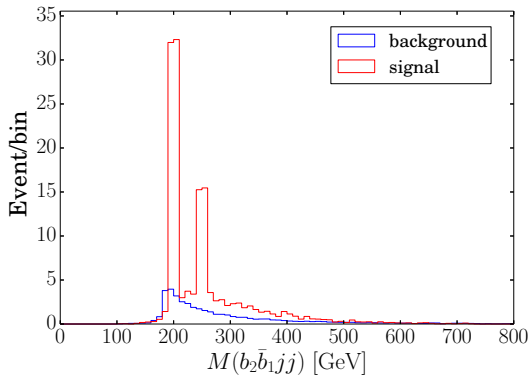}
\caption{The distribution of the invariant mass of $b_2 \overline{b}_1j j$ 
		in the signal and background events under the kinematical cuts 
		in Eq.~(\ref{eq:basic_cut}). 
		In the figure, the width of bins is $10~\mathrm{GeV}$. 
		We use the benchmark values in 
		Eq.(\ref{eq:benchmark_value_ScenarioII}).}
\label{fig:invariant_mass_bbjj}
\end{center}
\end{figure}

In Fig.~\ref{fig:invariant_mass_bbjj}, 
we show the distributions of signal and backgrounds 
for the invariant mass of $b_2 \overline{b}_1 j j$. 
The numbers of events at the twin peaks are $\mathcal{O}(30)$ and $\mathcal{O}(10)$, 
which are larger than thaose at the twin Jacobian peaks in the figure for $M_T(b_2 \overline{b}_1 jj)$ (the right figure of Fig~\ref{fig:signal_plot_ScenarioII}).
\begin{table}[h]
\begin{center}
\begin{tabular}{c|c|c|c|}
                   & Signal $S$ & Background $B$ & $S/\sqrt{S+B}$ \\ \hline
\begin{tabular}{c}
Basic cuts \\
(Eq.~(\ref{eq:basic_cut})) \\
\end{tabular}
		&   145 & 40 & 11 \\     \hline         
\end{tabular}
\caption{Numbers of signal event and background events under the basic cuts in Eq.~(\ref{eq:basic_cut}) in Scenario II. 
We assume that the efficiency of b-tagging is $70~\%$. 
We use the benchmark values in Eq.(\ref{eq:benchmark_value_ScenarioII}).}
\label{table:events_scenarioII}
\end{center}
\end{table}

Next, we discuss the background events at HL-LHC. 
We consider the process 
$p p \to t  \overline{t} b \overline{b} W^+ \to b b \overline{b}\overline{b} W^+ W^+ W^- \to b b \overline{b} \overline{b} \ell^+ \ell^{\prime +} \nu \nu j j$ as the background. 
As a result of the numerical calculation, 
$40$ events are expected to appear at HL-LHC as shown 
in Table.~\ref{table:events_scenarioII}.
This is the same order with the signal events. 
In Fig.~\ref{fig:signal_plot_ScenarioII}, 
the distributions of $M_T(b_1 \overline{b}_2 \ell^+ \ell^{\prime +} \cancel{E}_T)$ and $M_T(b_2 \overline{b}_1 j j )$ in the background events are shown. 
We use only the basic cuts in Eq.~(\ref{eq:basic_cut}) in the numerical calculation. 
Nevertheless, in the both figures of Fig.~\ref{fig:signal_plot_ScenarioII}, 
the number of signal events around the Jacobian peaks are much larger than thoes of the background events. 

In Fig.~\ref{fig:invariant_mass_bbjj}, 
the distribution of the background events 
for the invariant mass $M(b_2 \overline{b}_1 j j)$ in the background events 
are shown.  
The numbers of signal events around the two peaks are much larger than those of the background events. 

In summary, 
it would be possible that we obtain information on masses of 
all the charged scalars $H_1^\pm$, $H_2^\pm$, and $\Phi^{\pm\pm}$ 
by investigating the transverse mass distribution for $M_T(b_2 \overline{b}_1 \ell^+ \ell^{\prime +} \cancel{E}_T)$ and $M_T(b_1 \overline{b}_2 j j)$ and the invariant mass distribution for $M(b_1 \overline{b}_2 j j)$ at HL-LHC. 

Before closing Subsection B, 
we give a comment about the detector resolution. 
In the process of Scenario-II, 
the typical value of the transverse momenta of 
jets and bottom quarks is about $100~\mathrm{GeV}$. 
As mentioned in the end of the section for Scenario-I, 
at the ATLAS detector, 
the energy resolution for $p_T^j \simeq 100~\mathrm{GeV}$ 
is about $10~\%$~\cite{Aad:2020flx}. 
In Figs.~\ref{fig:signal_plot_ScenarioII} and \ref{fig:invariant_mass_bbjj}, 
we take the width of bins as $10~\mathrm{GeV}$. 
Therefore, 
it would be possible that the twin Jacobian peaks 
in the distribution for $M_T(jj)$ or $M(jj)$ overlap each other and 
they looks like one Jacobian peak with the unclear endpoint at the ATLAS detector if the mass differences is not large enough. 
Then, it would be difficult to obtain the information on both 
$m_{H_1}^{}$ and $m_{H_2}^{}$ from the transverse momentum 
distribution. 
Even in this case, 
it would be able to obtain the hint for masses by investigating the process.

\section{Summary and conclusion}

We have investigated collider signatures of the doubly and 
singly charged scalar bosons at the HL-LHC 
by looking at the transverse mass distribution as well as the invariant mass distribution 
in the minimal model with the isospin doublet with the hypercharge $Y=3/2$. 
We have discussed the background reduction for the signal process $pp \to W^{+\ast} \to \Phi^{++} H_{1,2}^-$ in the following two cases depending on the mass of the scalar bosons with the appropriate kinematical cuts . 
(1) The main decay mode of the singly charged scalar bosons is the tau lepton and missing (as well as charm and strange quarks). 
(2) That is into a top bottom pair.  
In the both cases, we have assumed that the doubly charged scalar boson is heavier than the singly charged ones. 
It has been concluded that the scalar doublet field with $Y=3/2$ is expected to be 
detectable for these cases at the HL-LHC unless the masses of $\Phi^{\pm\pm}$ and $H_{1,2}^\pm$ are too large.

\section*{Acknowledgements}
We would like to thank Arindam Das and Kei Yagyu for useful discussions.  
This work is supported by Japan Society for the 
Promotion of Science, Grant-in-Aid for Scientific Research, No. 16H06492, 18F18022, 18F18321 and 20H00160.

\newpage
\appendix

\section{Some formulae for the decays of charged scalars}

In this section, we show some analytic formulae for decay rates of the charged scalars $H_{1,2}^\pm$ and $\Phi^{\pm\pm}$. 

\subsection{Formulae for decays of the singly charged scalars $H_{1,2}^\pm$}
\label{sec:Decay_of_H}

\subsubsection{2-body decays}

The decay rate for the decay of $H_i^\pm$  $(i=1,2)$ into a pair of quarks $q q^\prime$ is given by 
\beq
\label{eq:Decay_H_to_qq}
\Gamma (H_i^\pm \to qq^\prime)
= \frac{ 3 m_{H_i}^{} }{ 8 \pi }
\left( \frac{ m_{H_i}^2 } { v^2 } \right)
\chi_i^{\prime 2} \cot^2 \beta |V_{q q^\prime}|^2 
\Bigl( (r_q + r_{q^\prime}) - (r_q + r_{q^\prime})^2 
- 4 r_q r_{q^\prime} \Bigr) 
F(r_q, r_{q^\prime}), 
\eeq
where $r_q$ ($r_{q^\prime}$) is the ratio of the squared mass of quark $q$ ($q^\prime$) 
to the squared mass of $H_i^\pm$: 
\beq
r_q = \frac{ m_q^2 }{ m_{H_i}^2 }, 
\quad
r_{q^\prime} = \frac{ m_{q^\prime}^2 }{ m_{H_i}^2 }, 
\eeq
and $\chi_i^\prime$ is defined as follows. 
\beq
\chi_1^\prime = \cos \chi, 
\quad
\chi_2^\prime = \sin \chi. 
\eeq
The function $F(x,y)$ in Eq.~(\ref{eq:Decay_H_to_qq}) is defined as 
\beq
\label{eq:F_function}
F(x,y) = \sqrt{ 1 + ( x - y )^2 - 2 ( x + y ) }.
\eeq
The decay rate for the decay of $H_i^\pm$ into a charged lepton $\ell$ and a neutrino $\nu_\ell$ is given by
\beq
\Gamma (H_i^\pm \to \ell \nu_\ell)
= \frac{ m_{H_i}^{} }{ 8 \pi }
\left( \frac{ m_\ell }{ v } \right)^2 
\chi_i^{\prime 2 } \cot^2 \beta 
\left( 
	1 - \frac{ m_\ell^2 }{ m_{H_i}^2 }
\right), 
\eeq
where $m_\ell$ is mass of $\ell$. 

In the case that $m_{H_i}^{} > m_{H_j}^{} + m_Z^{}$ $(i, j = 1,2, i \neq j)$, 
the decay $H_i^\pm \to H_j^\pm Z$ is allowed, and its decay rate is given by 
\beq
\Gamma(H_i\pm \to H_j^\pm Z)
= \frac{ m_{H_i} }{ 16 \pi }
\left( 
	\frac{ m_{H_i} }{ v } 
\right)^2 
\sin^2 2 \chi F ( r_Z, r_j )^3
\quad
(i\neq j), 
\eeq
where
\beq
\label{eq:rz_rj}
r_Z = \frac{ m_W^2 }{ m_{H_i}^2 }, 
\quad
r_{j} = \frac{ m_{H_j}^2 }{ m_{H_i}^2 }. 
\eeq

\subsubsection{3-body decays}

The decay rate for $H_i^\pm \to t^\ast b \to W^\pm b \overline{b}$ is given by
\beq
\Gamma(H_i^\pm \to t^\ast b \to W^\pm b \overline{b} )
= \frac{ 3 m_{H_i}^{} }{ 128 \pi^3 }
\left( 
	\frac{ m_t }{ v }
\right)^4
\chi_i^{\prime 2 }
\cot^2 \beta 
|V_{tb}|^2
\int_{r_W}^1 
\frac{ \mathrm{d}x }{ x }
\frac{ (1-x)^2 ( x - r_W^{} )^2 ( x + 2 r_W^{} ) }{ ( x - r_t )^2 + r_t r_{\Gamma_t}^{} },
\eeq
where mass of the bottom quark is neglected, and $r_W$, $r_t$, and $r_{\Gamma_r}^{}$ are defined as follows. 
\beq
r_W^{} = \frac{ m_W^2 }{ m_{H_i}^2 }, 
\quad
r_t = \frac{ m_t^2 }{ m_{H_i}^2 }, 
\quad
r_{\Gamma_t}^{} = \frac{ \Gamma_t^2 }{ m_{H_i}^2 }, 
\eeq
where $\Gamma_t$ is the total decay width of the top quark. 

In the case that $m_{H_i}^{} > m_{H_j}^{}$ ($i \neq j$), 
the decay $H_i^\pm \to H_j^\pm Z^\ast \to H_j^\pm f \overline{f}$, where $f$ is a SM fermion, 
is allowed. 
The decay rate is given by
\bal
\label{eq:Decay_H_to_Hff}
\Gamma ( H_i^\pm \to H_j^\pm Z^\ast \to H_j^\pm f \overline{f} )
= & \frac{ N_c^f m_{H_i}^{} }{ 192 \pi^3 }
\left( 
	\frac{ m_Z^{} }{ v }
\right)^4 
\sin^2 2 \chi 
\bigl( (C_V^f)^2 + (C_A^f)^2 \bigr)
\nonumber \\
& \times 
\int_0^{(1 - \sqrt{ r_j})^2}
\mathrm{d}x
\frac{ F(x, r_j)^3 }{ (x-r_Z^{})^2 + r_Z^{} r_{\Gamma_Z}^{} }, 
\eal
where $N_c^f$ is the color degree of freedom of a fermion $f$, 
$r_Z^{}$ and $r_j$ are defined same with that in Eq.~(\ref{eq:rz_rj}), 
and $r_{\Gamma_Z}^{}$ is the ratio of the squared decay rate of $Z$ boson to squared mass of $H_i^\pm$: 
\beq
r_{\Gamma_Z}^{}
= \frac{ \Gamma_Z^2 }{ m_{H_i}^2 }. 
\eeq
In addition, the coeffitient $C_V^f$ ($C_A^f$) in Eq.~(\ref{eq:Decay_H_to_Hff}) 
is the coupling constant of the vector (axial vector) current: 
\beq
\mathcal{L} = 
\frac{ g_L }{ 2 \cos \theta_W }
\overline{f} \gamma^\mu
\bigl( C_V^f + C_A^f \gamma_5 \bigr)
f Z_\mu, 
\eeq
where $g_L$ is the gauge coupling constant of the gauge group $SU(2)_L$ , 
and $\theta_W$ is the Weinberg angle. 
In Eq.~(\ref{eq:Decay_H_to_Hff}), mass of fermions are neglected. 

\subsection{Formulae for decays of the doubly charged scalar $\Phi^{\pm\pm}$}
\label{sec:Decay_of_Phi}

\subsubsection{2-body decay}

If $m_{\Phi^{\pm\mp}} > m_{H_i}^{} + m_W^{}$, 
the decay $\Phi^{\pm\pm} \to H_i^\pm W^\pm$ ($i=1,2$) is allowed. 
The decay rate is given by 
\beq
\Gamma ( \Phi^{\pm\pm} \to H_i^\pm W^\pm)
= \frac{ m_\Phi^{} }{ 8 \pi }
\left( 
	\frac{ m_\Phi  }{ v }
\right)^2
\chi_i^2 
F( R_W, R_i )^3, 
\eeq
where $\chi_i$ is defined in Eq.~(\ref{eq:chi_i}), 
the function $F(x,y)$ is defined in Eq.~(\ref{eq:F_function}), 
and $R_i$ and $R_W$ is defined as follows. 
\beq
R_W = \frac{ m_W^2 }{ m_\Phi^2 }, 
\quad
R_i = \frac{ m_{H_i}^2 }{ m_\Phi^2 }. 
\eeq

\subsubsection{3-body decay}

In the case that where the mass differences between $\Phi^{\pm\pm}$ and $H_i^\pm$ is so small that 
decays $\Phi^{\pm\pm} \to H_i^\pm W^\pm$ 
are prohibited,  
three-body decays $\Phi^{\pm\pm} \to H_i^\pm f \overline{f^\prime}$, where $f$ and $f^\prime$ are SM fermions, are dominant in small $m_{\Phi}^{}$ region. 
(See Fig.~\ref{fig:Decay_Phi}.) 
The branching ratio for $\Phi^{\pm \pm} \to H_i^\pm f \overline{f^\prime}$ 
is given by
\beq
\label{eq:Phi_to_Hff}
\Gamma ( \Phi^{\pm\pm} \to H_i^\pm f \overline{f^\prime})
= \frac{ N_c^f }{ 96 \pi^3 } \chi_i^2 
\int_0^{(1-\sqrt{R_i})^2 }
\frac{ \mathrm{d}x }{ x } \frac{ F(x, R_i)^3 }{ ( x - R_W )^2 + R_{\Gamma_W} R_W }, 
\eeq
where $R_{\Gamma_W}$ is the squared ratio of the decay width of $W$ boson ($\Gamma_W$) to $m_{\Phi}^{}$; 
\beq
R_{\Gamma_W} = \frac{ \Gamma_W^2 }{ m_{\Phi}^2 }.
\eeq
In Eq.~(\ref{eq:Phi_to_Hff}), 
we neglect the masses of $f$ and $f^\prime$.  

In the large $m_{\Phi}^{}$ region, 
$\Phi^{\pm\pm} \to W^\pm f \overline{f^\prime}$ is also important. 
The decay rate is given by
\bal
\Gamma(\Phi^{\pm\pm} \to W^\pm f \overline{f^\prime})
= &\frac{ N_c^f m_{\Phi}^{} }{ 256 \pi^3 }
\left( \frac{ m_{\Phi}^{} }{ v } \right)^4 
\sin 2 \chi \cot \beta^2 | V_{f f^\prime} |^2 
\nonumber \\[10pt]
& \times \int_{(\sqrt{R_f} + \sqrt{R_{f^\prime}})^2}^{(1-\sqrt{R_W})^2 }
\mathrm{d}x \ 
F\Bigl( \frac{ R_f }{ x }, \frac{ R_f^\prime }{ x } \Bigr)
F ( x, R_W ) G(x), 
\eal
where the function $G(x)$ is defined as follows. 
\bal
G(x) = &\Bigl\{ (R_f + R_{f^\prime}) ( x - R_f - R_{f^\prime}) - 4 R_f R_{f^\prime} \Bigr\}
\nonumber \\[10pt]
& \times \Biggl\{
\frac{ 1 }{ (x - R_1)^2 + R_1 R_{\Gamma_1} }
+ \frac{ 1 }{ (x - R_2)^2 + R_2 R_{\Gamma_2} }
\Biggr\}^2. 
\eal
The symbols $R_f$, $R_{f^\prime}$, $R_i$, and $R_{\Gamma_i}$ ($i=1,2$) are given by
\beq
R_f = \frac{ m_f^2 }{ m_{\Phi}^2 }, \quad
R_{f^\prime} = \frac{ m_{f^\prime}^2 }{ m_{\Phi}^2 }, \quad
R_i = \frac{ m_{H_i}^2 }{ m_{\Phi}^2 }, \quad
R_{\Gamma_i} = \frac{ \Gamma_{H_i}^2 }{ m_{\Phi}^2 }, 
\eeq
where $m_f$ ($m_{f^\prime}$) is mass of $f$ ($f^\prime$), 
and $\Gamma_{H_i}$ is the decay width of $H_i^\pm$.


\begin{thebibliography}{99}



\bibitem{ref:Higgs_discovery}
%
G.~Aad \textit{et al.} [ATLAS],
Phys. Lett. B \textbf{716}, 1-29 (2012)
[arXiv:1207.7214 [hep-ex]];
%
S.~Chatrchyan \textit{et al.} [CMS],
Phys. Lett. B \textbf{716}, 30-61 (2012)
[arXiv:1207.7235 [hep-ex]].


\bibitem{ref:Type-I_seesaw}
%
P.~Minkowski,
Phys. Lett. B \textbf{67}, 421-428 (1977); 
%
T.~Yanagida,
Conf. Proc. C \textbf{7902131}, 95-99 (1979); 
KEK-79-18-95; 
Prog. Theor. Phys. \textbf{64}, 1103 (1980); 
%
M.~Gell-Mann, P.~Ramond and R.~Slansky,
Conf. Proc. C \textbf{790927}, 315-321 (1979)
[arXiv:1306.4669 [hep-th]]; 
%
R.~N.~Mohapatra and G.~Senjanovic,
Phys. Rev. Lett. \textbf{44}, 912 (1980). 


\bibitem{ref:Type-II_seesaw}
%
W.~Konetschny and W.~Kummer,
Phys. Lett. B \textbf{70}, 433-435 (1977); 
%
M.~Magg and C.~Wetterich,
Phys. Lett. B \textbf{94}, 61-64 (1980); 
%
J.~Schechter and J.~W.~F.~Valle,
Phys. Rev. D \textbf{22}, 2227 (1980); 
%
G.~Lazarides, Q.~Shafi and C.~Wetterich,
Nucl. Phys. B \textbf{181}, 287-300 (1981). 


\bibitem{ref:Left-Right}
%
R.~N.~Mohapatra and G.~Senjanovic,
Phys. Rev. Lett. \textbf{44}, 912 (1980); 
%
Phys. Rev. D \textbf{23}, 165 (1981). 


\bibitem{ref:Type-III_seesaw}
R.~Foot, H.~Lew, X.~G.~He and G.~C.~Joshi,
Z. Phys. C \textbf{44}, 441 (1989)


\bibitem{ref:Zee}
A.~Zee,
Phys. Lett. B \textbf{93}, 389 (1980)
[erratum: Phys. Lett. B \textbf{95}, 461 (1980)]


\bibitem{ref:Zee_Babu}
%
A.~Zee,
Nucl. Phys. B \textbf{264}, 99-110 (1986); 
%
K.~S.~Babu,
Phys. Lett. B \textbf{203}, 132-136 (1988). 


\bibitem{ref:Cheng_Li}
T.~P.~Cheng and L.~F.~Li,
Phys. Rev. D \textbf{22}, 2860 (1980). 


\bibitem{ref:KNT}
L.~M.~Krauss, S.~Nasri and M.~Trodden,
Phys. Rev. D \textbf{67}, 085002 (2003)
[arXiv:hep-ph/0210389 [hep-ph]].


\bibitem{ref:Ma}
E.~Ma,
Phys. Rev. D \textbf{73}, 077301 (2006)
[arXiv:hep-ph/0601225 [hep-ph]].


\bibitem{ref:AKS}
%
M.~Aoki, S.~Kanemura and O.~Seto,
Phys. Rev. Lett. \textbf{102}, 051805 (2009)
[arXiv:0807.0361 [hep-ph]]; 
%
Phys. Rev. D \textbf{80}, 033007 (2009)
[arXiv:0904.3829 [hep-ph]]; 
%
M.~Aoki, S.~Kanemura and K.~Yagyu,
Phys. Rev. D \textbf{83}, 075016 (2011)
[arXiv:1102.3412 [hep-ph]].


\bibitem{ref:Cocktail}
M.~Gustafsson, J.~M.~No and M.~A.~Rivera,
Phys. Rev. Lett. \textbf{110}, no.21, 211802 (2013)
[erratum: Phys. Rev. Lett. \textbf{112}, no.25, 259902 (2014)]
[arXiv:1212.4806 [hep-ph]]; 
%
Phys. Rev. D \textbf{90}, no.1, 013012 (2014)
[arXiv:1402.0515 [hep-ph]].


\bibitem{Araki:2011hm}
T.~Araki, C.~Q.~Geng and K.~I.~Nagao,
Phys. Rev. D \textbf{83}, 075014 (2011)
[arXiv:1102.4906 [hep-ph]].


\bibitem{ref:Deshpande_Ma}
N.~G.~Deshpande and E.~Ma,
Phys. Rev. D \textbf{18}, 2574 (1978). 


\bibitem{ref:intert_singlet}
%
J.~McDonald,
Phys. Rev. D \textbf{50}, 3637-3649 (1994)
[arXiv:hep-ph/0702143 [hep-ph]]; 
%
C.~P.~Burgess, M.~Pospelov and T.~ter Veldhuis,
Nucl. Phys. B \textbf{619}, 709-728 (2001)
[arXiv:hep-ph/0011335 [hep-ph]]; 
%
S.~Kanemura, S.~Matsumoto, T.~Nabeshima and N.~Okada,
Phys. Rev. D \textbf{82}, 055026 (2010)
[arXiv:1005.5651 [hep-ph]].



\bibitem{ref:Kobayashi_Maskawa}
M.~Kobayashi and T.~Maskawa,
Prog. Theor. Phys. \textbf{49}, 652-657 (1973)

\bibitem{ref:Lee_CPviolation}
T.~D.~Lee,
Phys. Rev. D \textbf{8}, 1226-1239 (1973). 


\bibitem{Kuzmin:1985mm}
V.~A.~Kuzmin, V.~A.~Rubakov and M.~E.~Shaposhnikov,
Phys. Lett. B \textbf{155}, 36 (1985). 

\bibitem{Cohen:1990it}
A.~G.~Cohen, D.~B.~Kaplan and A.~E.~Nelson,
Nucl. Phys. B \textbf{349}, 727-742 (1991). 


\bibitem{Aoki:2011yk}
M.~Aoki, S.~Kanemura and K.~Yagyu,
Phys. Lett. B \textbf{702}, 355-358 (2011)
[erratum: Phys. Lett. B \textbf{706}, 495-495 (2012)]
[arXiv:1105.2075 [hep-ph]].

\bibitem{Okada:2015hia}
H.~Okada and K.~Yagyu,
Phys. Rev. D \textbf{93} (2016) no.1, 013004
[arXiv:1508.01046 [hep-ph]].

\bibitem{Cheung_Okada}
K.~Cheung and H.~Okada,
Phys. Lett. B \textbf{774} (2017), 446-450
[arXiv:1708.06111 [hep-ph]];
%

\bibitem{Enomoto:2019mzl}
K.~Enomoto, S.~Kanemura, K.~Sakurai and H.~Sugiyama,
Phys. Rev. D \textbf{100} (2019) no.1, 015044
[arXiv:1904.07039 [hep-ph]].

\bibitem{Ma:2019coj}
E.~Ma,
Phys. Lett. B \textbf{809} (2020), 135736
[arXiv:1912.11950 [hep-ph]].

\bibitem{Das:2020pai}
A.~Das, K.~Enomoto, S.~Kanemura and K.~Yagyu,
Phys. Rev. D \textbf{101} (2020) no.9, 095007
[arXiv:2003.05857 [hep-ph]].


\bibitem{Georgi:1985nv}
H.~Georgi and M.~Machacek,
Nucl. Phys. B \textbf{262}, 463-477 (1985)


\bibitem{ArkaniHamed:2002qy}
N.~Arkani-Hamed, A.~G.~Cohen, E.~Katz and A.~E.~Nelson,
JHEP \textbf{07}, 034 (2002)
[arXiv:hep-ph/0206021 [hep-ph]].


\bibitem{ref:Gunion}
J.~F.~Gunion,
Int. J. Mod. Phys. A \textbf{11} (1996), 1551-1562
[arXiv:hep-ph/9510350 [hep-ph]];

\bibitem{ref:triplet_pheno}
A.~G.~Akeroyd and M.~Aoki,
Phys. Rev. D \textbf{72}, 035011 (2005)
[arXiv:hep-ph/0506176 [hep-ph]]; 
%
A.~G.~Akeroyd, C.~W.~Chiang and N.~Gaur,
JHEP \textbf{11}, 005 (2010)
[arXiv:1009.2780 [hep-ph]]; 
%
A.~G.~Akeroyd and H.~Sugiyama,
Phys. Rev. D \textbf{84}, 035010 (2011)
[arXiv:1105.2209 [hep-ph]]; 
%
M.~Aoki, S.~Kanemura and K.~Yagyu,
Phys. Rev. D \textbf{85}, 055007 (2012)
[arXiv:1110.4625 [hep-ph]].

\bibitem{Han:2007bk}
T.~Han, B.~Mukhopadhyaya, Z.~Si and K.~Wang,
Phys. Rev. D \textbf{76}, 075013 (2007)
[arXiv:0706.0441 [hep-ph]].


\bibitem{Kanemura:2014goa}
S.~Kanemura, M.~Kikuchi, K.~Yagyu and H.~Yokoya,
Phys. Rev. D \textbf{90}, no.11, 115018 (2014)
[arXiv:1407.6547 [hep-ph]];
%
PTEP \textbf{2015}, 051B02 (2015)
[arXiv:1412.7603 [hep-ph]].

\bibitem{Rentala:2011mr}
V.~Rentala, W.~Shepherd and S.~Su,
Phys. Rev. D \textbf{84} (2011), 035004
[arXiv:1105.1379 [hep-ph]].

\bibitem{King:2014uha}
S.~F.~King, A.~Merle and L.~Panizzi,
JHEP \textbf{11}, 124 (2014)
[arXiv:1406.4137 [hep-ph]].

\bibitem{ref:distinguish_doubly}
H.~Sugiyama, K.~Tsumura and H.~Yokoya,
Phys. Lett. B \textbf{717}, 229-234 (2012)
[arXiv:1207.0179 [hep-ph]];
%
A.~Alloul, M.~Frank, B.~Fuks and M.~Rausch de Traubenberg,
Phys. Rev. D \textbf{88}, 075004 (2013)
[arXiv:1307.1711 [hep-ph]]; 
%
T.~Nomura, H.~Okada and H.~Yokoya,
Nucl. Phys. B \textbf{929}, 193-206 (2018)
[arXiv:1702.03396 [hep-ph]].

\bibitem{Vega:1989tt}
R.~Vega and D.~A.~Dicus,
Nucl. Phys. B \textbf{329}, 533-546 (1990)

\bibitem{Han:2003wu}
T.~Han, H.~E.~Logan, B.~McElrath and L.~T.~Wang,
Phys. Rev. D \textbf{67}, 095004 (2003)
[arXiv:hep-ph/0301040 [hep-ph]].

\bibitem{ref:exotic_Higgs}
S.~Kanemura, M.~Kikuchi and K.~Yagyu,
Phys. Rev. D \textbf{88}, 015020 (2013)
[arXiv:1301.7303 [hep-ph]]; 
%
J.~Hisano and K.~Tsumura,
Phys. Rev. D \textbf{87}, 053004 (2013)
[arXiv:1301.6455 [hep-ph]].

\bibitem{HL-LHC}
ATLAS collaboration, 
``Technical Design Report: A High-Granularity Timing Detector for the ATLAS Phase-II Upgrade", 
ATLAS-TDR-031 (2020); 
CMS collaboration, 
``The Phase-2 Upgrade of the CMS Level-1 Trigger", 
CMS-TDR-021 (2020). 



\bibitem{Glashow:1976nt}
S.~L.~Glashow and S.~Weinberg,
Phys. Rev. D \textbf{15}, 1958 (1977).

\bibitem{Branco:2011iw}
G.~C.~Branco, P.~M.~Ferreira, L.~Lavoura, M.~N.~Rebelo, M.~Sher and J.~P.~Silva,
Phys. Rept. \textbf{516}, 1-102 (2012)
[arXiv:1106.0034 [hep-ph]].

\bibitem{Aoki:2009ha}
M.~Aoki, S.~Kanemura, K.~Tsumura and K.~Yagyu,
Phys. Rev. D \textbf{80}, 015017 (2009)
[arXiv:0902.4665 [hep-ph]].

\bibitem{Cabibbo:1963yz}
N.~Cabibbo,
Phys. Rev. Lett. \textbf{10}, 531-533 (1963)

\bibitem{Enomoto:2015wbn}
T.~Enomoto and R.~Watanabe,
JHEP \textbf{05}, 002 (2016)
[arXiv:1511.05066 [hep-ph]].

\bibitem{Haller:2018nnx}
J.~Haller, A.~Hoecker, R.~Kogler, K.~M\"onig, T.~Peiffer and J.~Stelzer,
Eur. Phys. J. C \textbf{78}, no.8, 675 (2018)
[arXiv:1803.01853 [hep-ph]].

\bibitem{Arbey:2017gmh}
A.~Arbey, F.~Mahmoudi, O.~Stal and T.~Stefaniak,
Eur. Phys. J. C \textbf{78}, no.3, 182 (2018)
[arXiv:1706.07414 [hep-ph]].

\bibitem{Aiko:2020ksl}
M.~Aiko, S.~Kanemura, M.~Kikuchi, K.~Mawatari, K.~Sakurai and K.~Yagyu,
[arXiv:2010.15057 [hep-ph]].

\bibitem{Abbiendi:2013hk}
G.~Abbiendi \textit{et al.} [ALEPH, DELPHI, L3, OPAL and LEP],
Eur. Phys. J. C \textbf{73}, 2463 (2013)
[arXiv:1301.6065 [hep-ex]].



\bibitem{Ma:1997up}
E.~Ma, D.~P.~Roy and J.~Wudka,
Phys. Rev. Lett. \textbf{80}, 1162-1165 (1998)
[arXiv:hep-ph/9710447 [hep-ph]].

\bibitem{CapdequiPeyranere:1990qk}
M.~Capdequi Peyranere, H.~E.~Haber and P.~Irulegui,
Phys. Rev. D \textbf{44}, 191-201 (1991); 
%
S.~Kanemura,
Phys. Rev. D \textbf{61}, 095001 (2000)
[arXiv:hep-ph/9710237 [hep-ph]].

\bibitem{ref:pair_production}
J.~F.~Gunion and H.~E.~Haber,
Nucl. Phys. B \textbf{278}, 449 (1986)
[erratum: Nucl. Phys. B \textbf{402}, 569-569 (1993)]; 
%
S.~S.~D.~Willenbrock,
Phys. Rev. D \textbf{35}, 173 (1987); 
%
O.~Brein and W.~Hollik,
Eur. Phys. J. C \textbf{13}, 175-184 (2000)
[arXiv:hep-ph/9908529 [hep-ph]]; 
%
A.~A.~Barrientos Bendezu and B.~A.~Kniehl,
Phys. Rev. D \textbf{64}, 035006 (2001)
[arXiv:hep-ph/0103018 [hep-ph]].



\bibitem{ref:Associated_production}
S.~Kanemura and C.~P.~Yuan,
Phys. Lett. B \textbf{530}, 188-196 (2002)
[arXiv:hep-ph/0112165 [hep-ph]]; 
%
Q.~H.~Cao, S.~Kanemura and C.~P.~Yuan,
Phys. Rev. D \textbf{69}, 075008 (2004)
[arXiv:hep-ph/0311083 [hep-ph]]; 
%
A.~Belyaev, Q.~H.~Cao, D.~Nomura, K.~Tobe and C.~P.~Yuan,
Phys. Rev. Lett. \textbf{100}, 061801 (2008)
[arXiv:hep-ph/0609079 [hep-ph]].

\bibitem{Gunion:1986pe}
J.~F.~Gunion, H.~E.~Haber, F.~E.~Paige, W.~K.~Tung and S.~S.~D.~Willenbrock,
Nucl. Phys. B \textbf{294}, 621 (1987)

\bibitem{Moretti:1996ra}
S.~Moretti and K.~Odagiri,
Phys. Rev. D \textbf{55}, 5627-5635 (1997)
[arXiv:hep-ph/9611374 [hep-ph]].

\bibitem{ref:WH_associated}
D.~A.~Dicus, J.~L.~Hewett, C.~Kao and T.~G.~Rizzo,
Phys. Rev. D \textbf{40}, 787 (1989); 
%
A.~A.~Barrientos Bendezu and B.~A.~Kniehl,
Phys. Rev. D \textbf{59}, 015009 (1999)
[arXiv:hep-ph/9807480 [hep-ph]];
%
S.~Moretti and K.~Odagiri,
Phys. Rev. D \textbf{59}, 055008 (1999)
[arXiv:hep-ph/9809244 [hep-ph]].

\bibitem{Asakawa:2005nx}
E.~Asakawa, O.~Brein and S.~Kanemura,
Phys. Rev. D \textbf{72}, 055017 (2005)
[arXiv:hep-ph/0506249 [hep-ph]].

\bibitem{ref:gluon_fusion}
%
A.~A.~Barrientos Bendezu and B.~A.~Kniehl,
Phys. Rev. D \textbf{61}, 097701 (2000)
[arXiv:hep-ph/9909502 [hep-ph]];
%
O.~Brein, W.~Hollik and S.~Kanemura,
Phys. Rev. D \textbf{63}, 095001 (2001)
[arXiv:hep-ph/0008308 [hep-ph]].

\bibitem{Akeroyd:2016ymd}
A.~G.~Akeroyd, M.~Aoki, A.~Arhrib, L.~Basso, I.~F.~Ginzburg, R.~Guedes, J.~Hernandez-Sanchez, K.~Huitu, T.~Hurth and M.~Kadastik, \textit{et al.}
Eur. Phys. J. C \textbf{77}, no.5, 276 (2017)
[arXiv:1607.01320 [hep-ph]].



\bibitem{Alwall:2014hca}
J.~Alwall, R.~Frederix, S.~Frixione, V.~Hirschi, F.~Maltoni, O.~Mattelaer, H.~S.~Shao, T.~Stelzer, P.~Torrielli and M.~Zaro,
JHEP \textbf{07} (2014), 079
[arXiv:1405.0301 [hep-ph]].

\bibitem{FeynRules}
N.~D.~Christensen and C.~Duhr,
Comput. Phys. Commun. \textbf{180} (2009), 1614-1641
[arXiv:0806.4194 [hep-ph]];
%
A.~Alloul, N.~D.~Christensen, C.~Degrande, C.~Duhr and B.~Fuks,
Comput. Phys. Commun. \textbf{185} (2014), 2250-2300
[arXiv:1310.1921 [hep-ph]].

\bibitem{Zyla:2020zbs}
P.~A.~Zyla \textit{et al.} [Particle Data Group],
PTEP \textbf{2020}, no.8, 083C01 (2020)

\bibitem{Sirunyan:2018pgf}
A.~M.~Sirunyan \textit{et al.} [CMS],
JINST \textbf{13}, no.10, P10005 (2018)
[arXiv:1809.02816 [hep-ex]].

\bibitem{Hagiwara:2012vz}
K.~Hagiwara, T.~Li, K.~Mawatari and J.~Nakamura,
Eur. Phys. J. C \textbf{73} (2013), 2489
[arXiv:1212.6247 [hep-ph]].

\bibitem{Ballestrero:2018anz}
A.~Ballestrero, B.~Biedermann, S.~Brass, A.~Denner, S.~Dittmaier, R.~Frederix, P.~Govoni, M.~Grossi, B.~J\"ager and A.~Karlberg, \textit{et al.}
Eur. Phys. J. C \textbf{78}, no.8, 671 (2018)
[arXiv:1803.07943 [hep-ph]].

\bibitem{Aad:2020flx}
G.~Aad \textit{et al.} [ATLAS],
[arXiv:2007.02645 [hep-ex]].

\bibitem{Sirunyan:2017ezt}
A.~M.~Sirunyan \textit{et al.} [CMS],
JINST \textbf{13}, no.05, P05011 (2018)
[arXiv:1712.07158 [physics.ins-det]].





\end{thebibliography}
\end{document}